\documentclass[usletter,11pt]{article}

\usepackage{jheppub} 
\usepackage{amsmath,amssymb}
\usepackage{graphicx}
\usepackage{subcaption}
\usepackage[percent]{overpic}
\usepackage{cancel}

\def\be{\begin{equation}}
\def\ee{\end{equation}}
\def\bea{\begin{eqnarray}}
\def\eea{\end{eqnarray}}

\newcommand\bZ{\mathbb{Z}}

\newcommand\cN{\mathcal{N}}

\newcommand\cB{\mathcal{B}}
\newcommand\cD{\mathcal{D}}
\newcommand\cC{\mathcal{C}}

\newcommand\cH{\mathcal{H}}

\newcommand\cQ{\mathcal{Q}}

\newcommand\cS{\mathcal{S}}

\newcommand\ex{\mathrm{e}}
\newcommand\ii{\mathrm{i}}
\newcommand\Tr{\mathrm{Tr}}
\newcommand\wri{\mathrm{wr}}

\newcommand\qqq{\qquad\qquad}
\newcommand{\nn}{\nonumber \\ {} }

\title{BPS spectra from BPS graphs}

\author{Maxime Gabella}

\affiliation{Institute for Advanced Study, Einstein Drive, Princeton, NJ 08540, USA}
\affiliation{Scuola Internazionale Superiore di Studi Avanzati, Via Bonomea 265, 34136, Trieste, Italy}

\emailAdd{maxime.gabella@gmail.com}

\abstract{
I present a simple graphical method to find the BPS spectra of $A_1$ theories of class~S. BPS graphs provide a bridge between spectral networks and BPS quivers, the two main frameworks for the study of BPS states. Here I show how to essentially read off from a BPS graph
the quantum spectrum generator (or BPS monodromy), expressed as a product of quantum dilogarithms.
Thanks to the framed wall-crossing phenomenon for line defects, the determination of the BPS spectrum reduces to the computation of quantum parallel transport across the edges of the BPS graph. 

\

}

\dedicated{To Charlotte}

\begin{document} 

\maketitle
\flushbottom

\begin{quote}
\begin{center}
\textit{``Considerate la vostra semenza:\\
fatti non foste a viver come bruti,\\
ma per seguir virtute e canoscenza.''}%
\end{center}
\begin{flushright}
{\sc Dante}, \emph{Inferno}, Canto XXVI: Ulysses (1320)
\end{flushright}
\end{quote}

\section{Introduction}

Many insights into non-perturbative aspects of four-dimensional gauge theories with $\cN=2$ supersymmetry have been gained from the analysis of their BPS states, which form a protected sector of the Hilbert space. 
BPS states are defined on the Coulomb branch $\cB$, where the gauge symmetry is broken to $U(1)^r$, and saturate the bound $M \geq |Z_\gamma|$, for the mass $M$,   charge~$\gamma$, and $\cN=2$ central charge $Z_\gamma$.
The two main frameworks for studying the spectrum of BPS states are spectral networks~\cite{Gaiotto:2009hg, Gaiotto:2012rg, Gaiotto:2012db} and BPS quivers~\cite{Cecotti:2010fi, Cecotti:2011rv, Cecotti:2011gu, Alim:2011ae, DelZotto:2011an, Alim:2011kw}. 
Spectral networks can detect BPS states as certain degenerate trajectories on the Riemann surface $\cC$ associated with a theory of class~S~\cite{Gaiotto:2009hg, Gaiotto:2009we}, while in BPS quivers BPS states correspond to stable quiver representations. 
In~\cite{Gabella:2017hpz}, we introduced the concept of \emph{BPS graph} to bridge the gap between these two approaches. A BPS graph is a very special type of spectral network that appears at the maximal intersection of walls of marginal stability on the Coulomb branch $\cB$, where the central charges of all the BPS states have the same phase. On the other hand, the topology of a BPS graph naturally encodes a BPS quiver: edges correspond to nodes and intersections to arrows. 

In this paper I show that BPS graphs also provide a simple way to derive the BPS spectrum. 
I take advantage of the interaction between ordinary BPS states and \emph{framed} BPS states, which appear in the presence of line defects. The phenomenon of framed wall-crossing~\cite{Gaiotto:2010be} indeed implies that the generating function $F(\wp,\vartheta)$ of framed BPS degeneracies transforms by conjugation by the \emph{quantum spectrum generator} $\cS$, aka the BPS monodromy:
\be\label{FpSFmS}
F(\wp,\vartheta+\epsilon) =  \cS F(\wp,\vartheta-\epsilon )  \cS^{-1} .
\ee
Given that $F(\wp,\vartheta)$ can be computed as a quantum holonomy along the path $\wp$ on $\cC$~\cite{Galakhov:2014xba, Gabella:2016zxu}, the framed wall-crossing formula can be turned around and used to determine the quantum spectrum generator $\cS$. 
The method consists in choosing a path $\wp$ that crosses an edge $\gamma_\alpha$ of the BPS graph and computing $F(\wp,\vartheta)$, or rather some building block $\cQ_\alpha$, expressed as a product of quantum dilogarithms. This only provides some partial information about $\cS$, but the procedure can be repeated for another edge $\gamma_\beta$ after having deleted the edge $\gamma_\alpha$ from the BPS graph. 
A few more iterations eventually produce the full quantum spectrum generator~$\cS$, with the schematic form
\be
\cS = \cQ_\alpha \cQ_{\beta, \cancel\alpha} \cdots = \prod^{\curvearrowleft}_{\gamma\in \Gamma} \Phi(X_\gamma).
\ee
Each factor $\Phi(X_\gamma)$ in this product correspond to a BPS state of charge $\gamma$ in the spectrum. The ordering follows from the choice of sequence of edges $\gamma_\alpha, \gamma_\beta, \ldots$, and reflects the ordering of the phases of the central charges $Z_\gamma$. Changing this sequence leads to the BPS spectrum in a different chamber on the Coulomb branch. However, since $\cS$ is a wall-crossing invariant, all choices are equivalent via the wall-crossing formula (quantum pentagon identity).

Here I will focus on $A_1$ theories of class~S but it should be possible to extend our method to determine the BPS spectra of higher-rank $A_{N-1}$ theories, whose BPS graphs were presented in~\cite{Gabella:2017hpz}.

\section{Spectral networks and BPS graphs}

\begin{figure}[tb]
\centering
\begin{minipage}{.32\textwidth}
\centering
	\begin{overpic}[width=\textwidth]{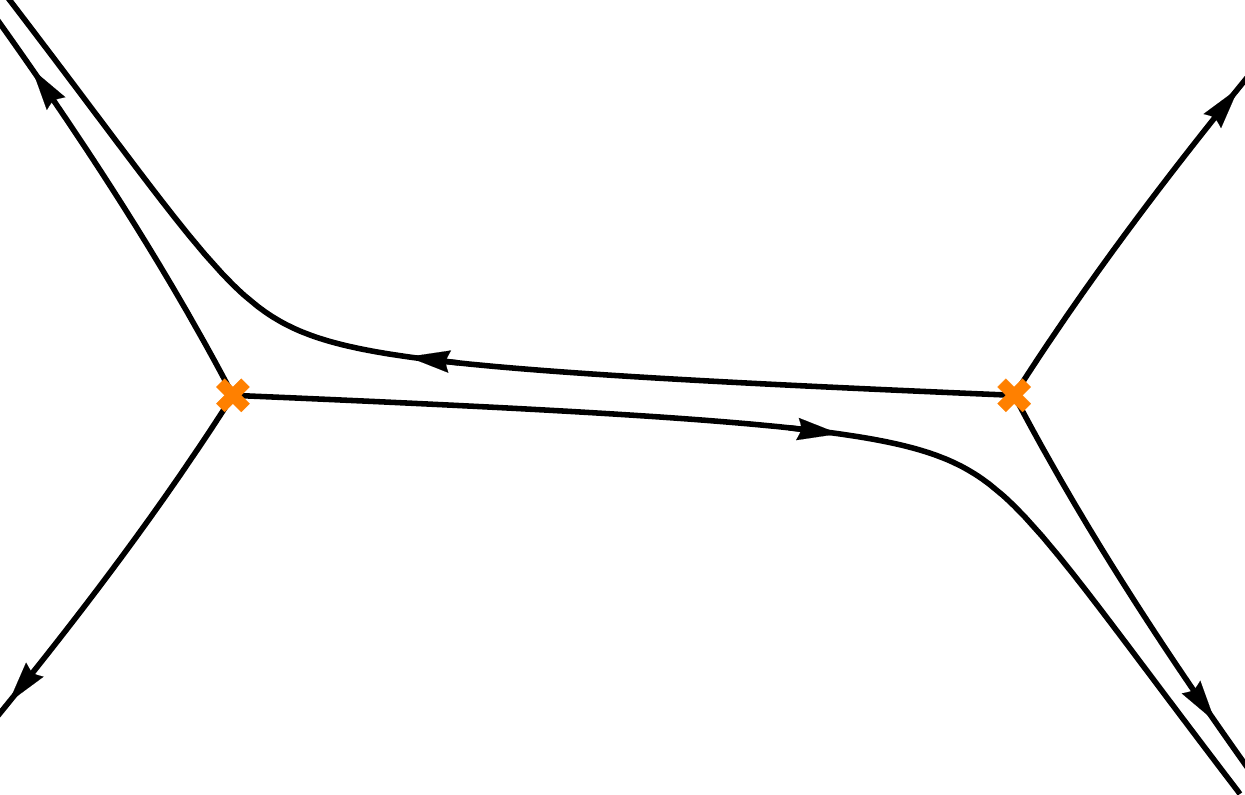}
 		\put (45,66) {$\vartheta_c^- $}
	\end{overpic}
\end{minipage} \hfill
\begin{minipage}{.32\textwidth}
\centering
	\begin{overpic}[width=\textwidth]{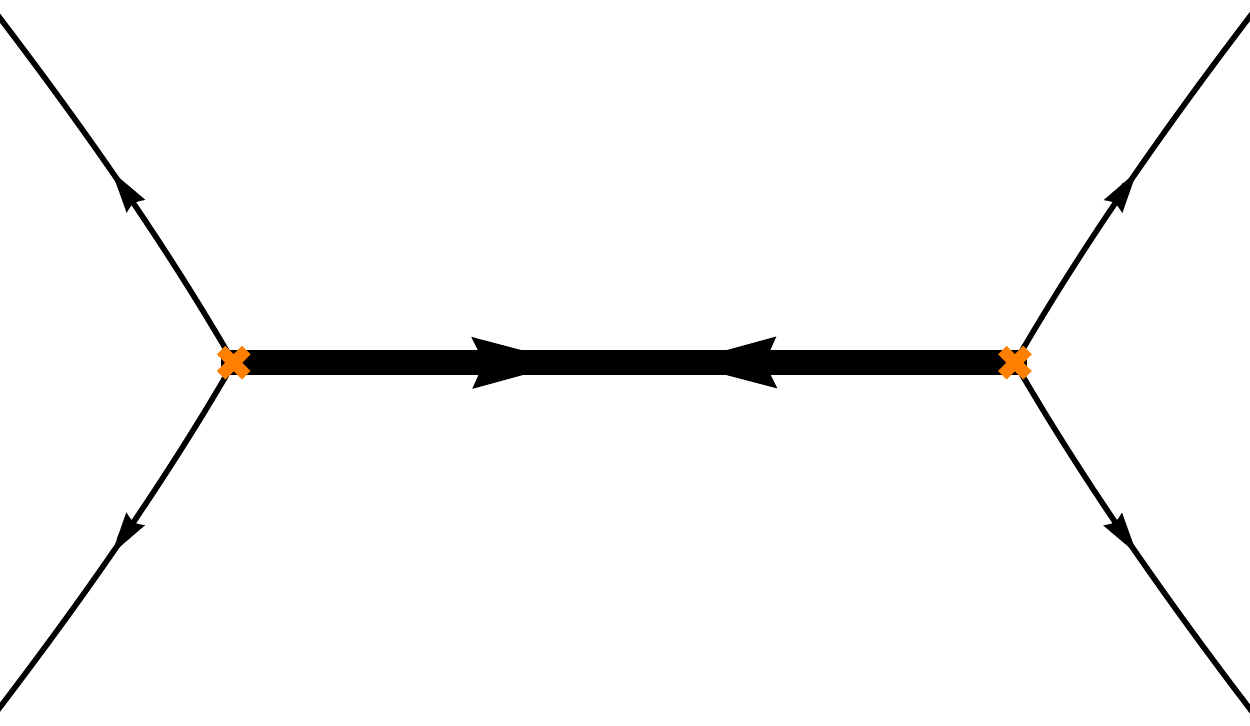}
 		\put (45,66) {$\vartheta_c$}
 		\put (46, 36) {$\gamma$}
	\end{overpic}
\end{minipage}\hfill
\begin{minipage}{.32\textwidth}
\centering
	\begin{overpic}[width=\textwidth]{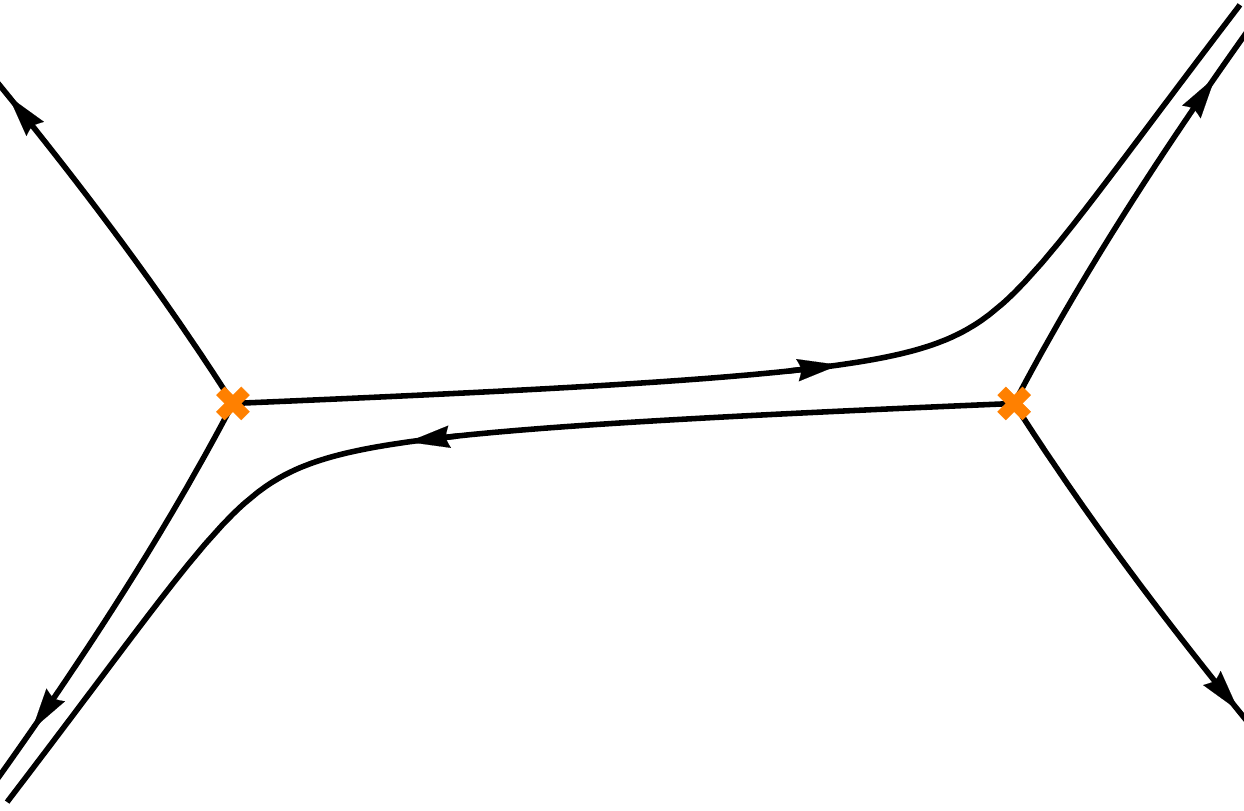}
 		\put (45,66) {$\vartheta_c^+ $}
	\end{overpic}
\end{minipage}
\caption{Appearance of a double wall in a spectral network for the critical phase $\vartheta = \vartheta_c$. \emph{Middle}: At $\vartheta_c$ two oppositely oriented walls merge into a double wall $\gamma$. \emph{Left}: American resolution at $\vartheta_c^- = \vartheta_c-\epsilon $. \emph{Right}: British resolution at $\vartheta_c^+= \vartheta_c+\epsilon $. }
\label{doublewall}
\end{figure}

Spectral networks~\cite{Gaiotto:2012rg, Gaiotto:2012db} are powerful geometric tools for the study of BPS states in theories of class~S. Recall that class~S comprises four-dimensional $\cN=2$ supersymmetric gauge theories obtained by compactification of the six-dimensional $(2,0)$ superconformal theory on a punctured Riemann surface $\cC$. Given an $A_{N-1}$ theory of class~S at a given point of its Coulomb branch~$\cB$, a spectral network on~$\cC$ is a collection of trajectories, or \emph{walls}, associated with the Seiberg-Witten curve $\Sigma$, an $N$-fold branched cover of $\cC$ (the spectral curve of the related Hitchin system). A wall typically starts at a branch point of the covering $\Sigma\to \cC$ and ends at a puncture (for $N>2$ there can also be joints where several walls meet). Simple branch points have three walls starting from them.
A spectral network also depends on a phase $\vartheta$, and its walls rotate around each branch point as $\vartheta$ varies. For critical values $\vartheta=\vartheta_c$, several walls can merge together and form a \emph{finite web} with all endpoints on branch points. The example of a \emph{double wall} is shown in Figure~\ref{doublewall}. A finite web lifts to a closed loop $\gamma$ on the cover $\Sigma$, which is naturally identified in the infrared abelian gauge theory with the electromagnetic and flavor charge $\gamma \in \Gamma = H_1(\Sigma,\bZ)$ of a BPS state with central charge $Z_\gamma = M \ex^{\ii\vartheta_c}$. 
The full BPS spectrum can thus in principle be obtained by varying the phase~$\vartheta$ from $0$ to $2\pi$ and recording all the double walls that appear. This approach can however become impossibly tedious beyond the simplest theories. 

The idea behind the concept of BPS graph~\cite{Gabella:2017hpz} is to have all the double walls appear simultaneously at a single critical phase $\vartheta_c$. This implies choosing a point of the Coulomb branch $\cB$ that is a maximal intersection of all walls of marginal stability, where the central charges of all BPS states have the same phase $\vartheta_c$. 
A BPS graph consists of all the finite webs that appear in the critical spectral network. 
For $A_1$ theories, its edges are double walls, and its vertices are branch points. 
The edges of a BPS graph provide a basis for the lattice $\Gamma$ of gauge and flavor charges of BPS states. 
The intersection form $\langle \gamma, \gamma' \rangle$ for two adjacent edges is given by $+1$ if $\gamma$ and $\gamma'$ are ordered clockwise around their common branch point (if they share more than one branch point we sum over them).
The topology of a BPS graph therefore encodes a BPS quiver, whose nodes correspond to edges and arrows to intersections.

\section{Framed wall-crossing}

Let's consider a theory of class~S associated with a Riemann surface $\cC$, and add a line defect~$L_{\wp,\vartheta}$ along the time direction.
The electromagnetic charge of $L_{\wp,\vartheta}$ is encoded by a path $\wp$ on $\cC$ and the supercharges that it preserves are specified by a phase $\vartheta$~\cite{Drukker:2009tz, Gaiotto:2010be}. 
The presence of $L_{\wp,\vartheta}$ modifies the Hilbert space and allows for a new type of BPS states, called \emph{framed BPS states}.
The framed BPS Hilbert space $\cH_{L_{\wp,\vartheta}}^\text{BPS}$ is graded by charges $\gamma\in \Gamma$:
\be
\cH_{L_{\wp,\vartheta}}^\text{BPS} = \bigoplus_{\gamma\in \Gamma} \cH_{L_{\wp,\vartheta},\gamma}^\text{BPS}.
\ee
Framed BPS states with charge $\gamma$ are counted (with spin) by the framed protected spin character~\cite{Gaiotto:2010be}:
\be
\underline{\overline\Omega}(L_{\wp,\vartheta} ,\gamma; q) = \Tr_{\cH_{L_{\wp,\vartheta},\gamma}^\text{BPS}} q^{J_3} (-q^\frac12)^{2I_3} ,
\ee
with $J_3$ and $I_3$ Cartan generators of $so(3)$ and $su(2)_R$.
These framed protected spin characters can be collected in the generating function
\be\label{genFuncF}
F(L_{\wp,\vartheta} ;q) = \sum_\gamma \underline{\overline\Omega}(L_{\wp,\vartheta} ,\gamma;q) X_\gamma,
\ee
where the noncommutative variables $X_\gamma$ satisfy
\be
X_\gamma X_{\gamma'} = q^{\frac12\langle \gamma,\gamma'\rangle} X_{\gamma+\gamma'}.
\ee
Here $\langle \cdot , \cdot \rangle : \Gamma\times \Gamma \to \bZ$ is the Dirac-Schwinger-Zwanziger antisymmetric product of charges, naturally identified with the intersection form on $H_1(\Sigma, \bZ)$. 
The expression~\eqref{genFuncF} specifies how the ultraviolet line defect $L_{\wp,\vartheta}$ decomposes into abelian line defects $X_\gamma$ in the infrared. 

The framed protected spin character $\underline{\overline\Omega}(L_{\wp,\vartheta} ,\gamma; q)$ can jump when the central charge $Z_\gamma = |Z_\gamma|\ex^{\ii\vartheta_c}$ of an ordinary BPS state aligns with the central charge of the line defect $L_{\wp,\vartheta}$. 
This is the \emph{framed wall-crossing} phenomenon.
Let's consider a path $\wp$ that crosses a double wall $\gamma$, and denote by $F^\pm(\wp)$ the generating functions $F(L_{\wp,\vartheta} ;q)$ for $\vartheta =\vartheta_c^\pm = \lim_{\epsilon \to 0^+} \vartheta_c \pm \epsilon$.
The framed wall-crossing formula can be elegantly expressed as a conjugation~\cite{Gaiotto:2010be}:
\be
F^+(\wp) =  \Phi(X_\gamma) F^-(\wp)  \Phi(X_\gamma)^{-1} .
\ee
Here $\Phi(X)$ is the quantum dilogarithm~\cite{1994MPLA....9..427F} defined as
\be
\Phi(X) = \prod_{n=0}^\infty \left( 1+ q^{n+\frac12} X\right)^{-1} .
\ee
It is uniquely characterized by the $q$-difference equations
\be
\Phi( q X) = (1+ q^{\frac12} X ) \Phi(X), \qqq \Phi( q^{-1} X) = (1+ q^{-\frac12} X )^{-1} \Phi(X),
\ee
and satisfies the quantum pentagon identity, which captures the simplest wall-crossing process where one hypermultiplet appears/disappears across a wall~\cite{Kontsevich:2008fj,Gaiotto:2008cd,Dimofte:2009bv,Dimofte:2009tm}:
\be\label{qPentagon}
\Phi(X_{\gamma_1})\Phi(X_{\gamma_2})=\Phi(X_{\gamma_2})\Phi(X_{\gamma_1+\gamma_2})\Phi(X_{\gamma_1}),
\ee
for $\langle \gamma_1, \gamma_2\rangle = 1$.

\newpage

\section{Quantum spectrum generator}

The particularity of the locus on the Coulomb branch $\cB$ for which a BPS graph appears is that all the BPS states have central charges with the same phase $\vartheta_c$. This implies that given a path $\wp$ that crosses an edge of a BPS graph, the framed wall-crossing formula takes the form
\be\label{FpSFmS}
F^+(\wp) =  \cS F^-(\wp)  \cS^{-1} ,
\ee
where $\cS$ is a product of quantum dilogarithms associated with all BPS states~\cite{Gaiotto:2010be} (see also~\cite{Longhi:2016wtv}).
The product $\cS$ is known as the \emph{quantum spectrum generator}, or the \emph{BPS monodromy}~\cite{Kontsevich:2008fj,Gaiotto:2008cd}, and is defined schematically as
\be
\cS = \prod_{\gamma\in \Gamma}^{\curvearrowleft}  \Phi( X_\gamma) ,
\ee
with the phase $\vartheta\in [0,\pi)$ increasing from right to left in the product (making sense of this ordering requires moving slightly away from the walls of marginal stability into a well-defined BPS chamber, which as we will see corresponds to choosing a sequence of edges in the BPS graph).
Now since the generating functions $F^\pm(\wp)$ can be computed as parallel transports with spin~\cite{Galakhov:2014xba}, or a quantum holonomies~\cite{Gabella:2016zxu}, the idea is to use the framed wall-crossing formula~\eqref{FpSFmS} to determine the quantum spectrum generator $\cS$. 

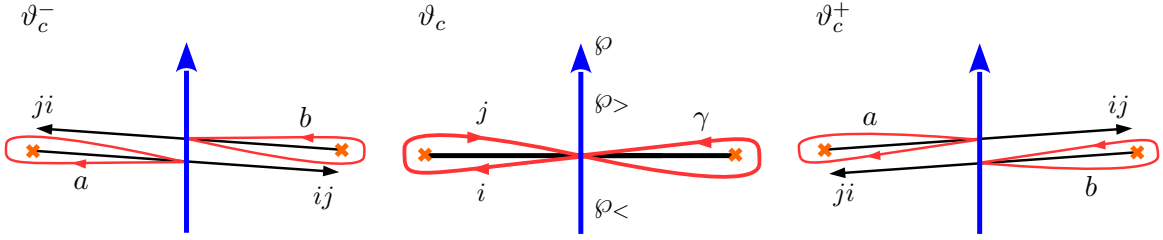
\begin{figure}[tb]
\begin{center}
\begin{overpic}[width=\textwidth]{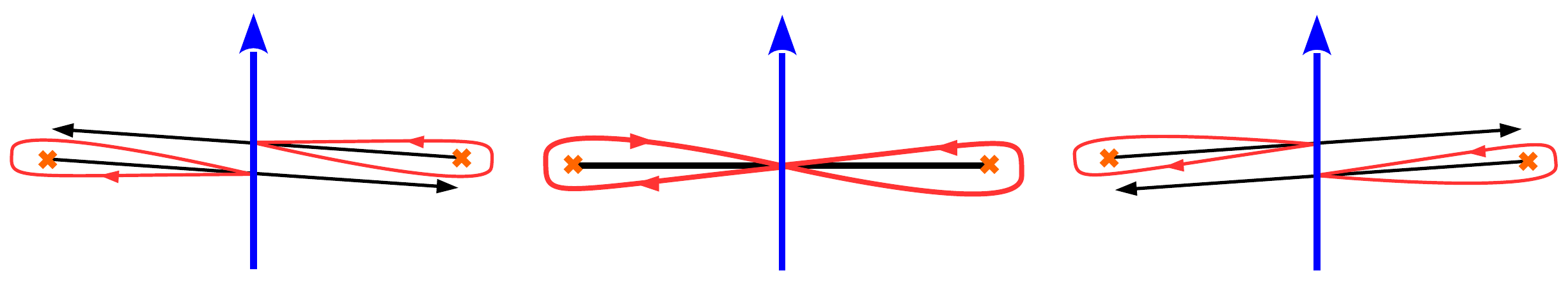}
 	\put (2, 19) {$\vartheta_c^-$}
 	\put (36, 19) {$\vartheta_c$}
 	\put (70, 19) {$\vartheta_c^+$}
 	\put (27, 3.7) {\small$ij$}
 	\put (3, 11.5) {\small$ji$}
 	\put (51, 17) {$\wp$}
 	\put (51, 3) {$\wp_<$}
 	\put (51, 12) {$\wp_>$}
 	\put (6.5, 5) {$a$}
 	\put (25.8, 10.3) {$b$}
 	\put (41, 4) {\small$i$} 	
	\put (41, 11) {\small$j$}
	\put (59.5, 10.5) {$\gamma$} 
	 \put (95, 11.5) {\small$ij$}
 	\put (71.5, 4) {\small$ji$}
 	\put (74, 10.5) {$a$}
 	\put (93, 4.4) {$b$}
\end{overpic}
\caption{Path $\wp$ across an edge $\gamma$ of a BPS graph. On the cover $\Sigma$, the closed loop $\gamma$ runs in one direction on sheet $i$ and in the other direction on sheet $j$. \emph{Left}: American resolution at $\vartheta_c^-$. \emph{Right}: British resolution at $\vartheta_c^+$. The detours $a\in \Gamma_{ij}$ and $b\in \Gamma_{ji}$ combine to give $\gamma$.}
\label{doubleWallPath}
\end{center}
\end{figure}

Let's consider a path $\wp$ that crosses an edge $\gamma$ of a BPS graph (Figure~\ref{doubleWallPath}).
In the American (at $\vartheta_c^-$) and British (at $\vartheta_c^+$) resolutions of the double wall $\gamma$, the generating functions $F^\pm(\wp)$ can be expressed schematically as
\bea
F^-(\wp) &=&  \cD(\wp_{<}) \Big( 1 + \sum_{a\in \Gamma_{ij}} X_a \Big)  \Big( 1 + \sum_{b\in \Gamma_{ji}} X_b \Big) \cD(\wp_{>}) , \\
F^+(\wp) &=&  \cD(\wp_{<})  \Big( 1 + \sum_{b\in \Gamma_{ji}} X_b \Big)\Big( 1 + \sum_{a\in \Gamma_{ij}} X_a \Big)  \cD(\wp_{>}) ,
\eea
where $\wp_<$ and $\wp_>$ are the halves of $\wp$ before and after the intersection with $\gamma$ and
\be
\cD (\wp) = \sum_{i=1}^N X_{\wp^{(i)}},
\ee
with $\wp^{(i)}$ the lift of $\wp$ to the $i$th sheet of $\Sigma$.
The paths $a$ and $b$ are detours along walls with label $ij$ and $ji$ respectively, as illustrated for a simple case in Figure~\ref{doubleWallPath} (more generally, detours can extend further along the edges of the BPS graph). 
It is convenient to define the power series
\bea\label{Qpmdef}
Q^-_\gamma &=& 1+ \sum_{a\in \Gamma_{ij}} \sum_{b\in \Gamma_{ji}}  q^{\frac12 \wri(ab) - \frac12 \langle \wp^{(i)}, ab \rangle } X_{ab} , \nn
Q^+_\gamma &=& 1+  \sum_{b\in \Gamma_{ji}} \sum_{a\in \Gamma_{ij}} q^{\frac12 \wri(ba) + \frac12 \langle \wp^{(j)}, ba \rangle } X_{ba} ,
\eea
where the writhe $\wri(ab)$ is the sum over self-intersections of the path $ab$. 
The conventions here are that an intersection with the right arm on top of the left arm gives $+1$, and that $\gamma'$ is on top of $\gamma$ in $\langle \gamma,\gamma'\rangle$.
The $ii$- and $jj$-components of $F^\pm$ then take the form
\bea
F^+_{ii} &=& X_{\wp^{(i)}}  , \qqq
F^-_{ii} =  X_{\wp^{(i)}} + \sum_{a\in \Gamma_{ij}}  \sum_{b\in \Gamma_{ji}} q^{\frac12 \wri(ab)  } X_{\wp_<^{(i)} ab \wp_>^{(i)}}   =
 X_{\wp^{(i)}}Q^-_p , \nn
F^+_{jj} &=& X_{\wp^{(j)}} +\sum_{b\in \Gamma_{ji}}  \sum_{a\in \Gamma_{ij}}  q^{\frac12 \wri(ba)  } X_{\wp_<^{(j)} ba \wp_>^{(j)}}  = Q^+_p  X_{\wp^{(j)}}  , \qqq
F^-_{jj} =  X_{\wp^{(j)}} ,
\eea
Commuting $\cS$ with $X_{\wp^{(i)}}$ or $X_{\wp^{(j)}}$ multiplies $X_\gamma$ in $\cS$ by $q^{\pm1}$ because $\langle \wp^{(i)}, \gamma \rangle = +1$ and $\langle \wp^{(j)}, \gamma \rangle = -1$.
With the notation
\be\label{notationShift}
\cS_{\gamma^\pm} = \cS|_{X_\gamma\to q^{\pm 1} X_\gamma} ,
\ee
the framed wall-crossing formula~\eqref{FpSFmS} finally implies
\bea\label{QpmSS}
Q^-_\gamma &=&  \cS_{\gamma^-}^{-1} \cdot \cS , \nn
Q^+_\gamma &=& \cS\cdot \cS_{\gamma^-}^{-1}.
\eea

As a simple check, let's take a BPS graph with a single edge.
In this case, there are only two possible combinations of detours, corresponding to $X_{ab} = X_{ba} = X_\gamma$, which gives
\be
Q_\gamma^- = Q_\gamma^+ = 1 + q^{-\frac12} X_{\gamma}.
\ee
On the other hand, the quantum generating function is simply $\cS = \Phi(X_\gamma)$ and so
\be
\cS_{\gamma^-}^{-1}  \cS=\cS \cS_{\gamma^-}^{-1} = \prod_{n=0}^\infty \left( 1+ q^{n-\frac12} X_\gamma\right) \left( 1+ q^{n+\frac12} X_\gamma\right)^{-1} = 1 + q^{-\frac12} X_{\gamma},
\ee
in agreement with~\eqref{QpmSS}.

\begin{figure}[tb]
\begin{center}
\begin{overpic}[width=\textwidth]{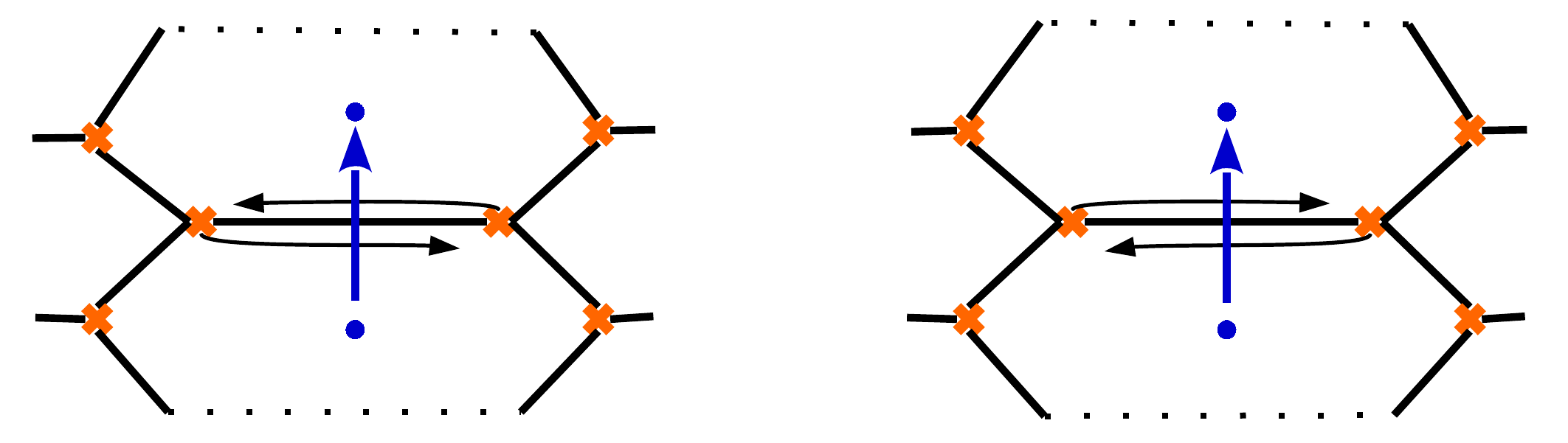}
\small
 	\put (25, 13.4) {$0$}
 	\put (7.5, 12) {$1$}
 	\put (7, 3) {$2$}
 	\put (3, 24) {$\bar m\!-\!1$}
 	\put (6, 15) {$\bar m$} 	
	\put (36, 12) {$m$}
 	\put (36, 3) {$m\!-\!1$}
 	\put (37, 24) {$\bar 2$}
 	\put (37, 15) {$\bar 1$}
 	\put (28, 10) {\small$ij$}
 	\put (15, 17) {\small$ji$}
	\put (81, 13.4) {$0$}
 	\put (63, 12) {$1$}
 	\put (62, 3) {$2$}
 	\put (58, 24) {$\bar m\!-\!1$}
 	\put (62, 15) {$\bar m$} 	
	\put (92, 12) {$m$}
 	\put (92, 3) {$m\!-\!1$}
 	\put (92.5, 24) {$\bar 2$}
 	\put (92, 15) {$\bar 1$}
 	\put (70, 9.5) {\small$ji$}
 	\put (84, 17) {\small$ij$}
\end{overpic}
\caption{Edge $\gamma_0$ in a generic BPS graph, at the interface of two cycles, $\gamma_0, \gamma_1, \gamma_2, \ldots, \gamma_m$ and $\gamma_0, \gamma_{\bar 1}, \gamma_{\bar 2}, \ldots, \gamma_{\bar m}$. \emph{Left}: American resolution at $ \vartheta_c^-$, in which the detours $a,b$ rotate anti-clockwise around the punctures. \emph{Right}: British resolution at $ \vartheta_c^+$, in which the detours  rotate clockwise. }
\label{Qgeneral}
\end{center}
\end{figure}

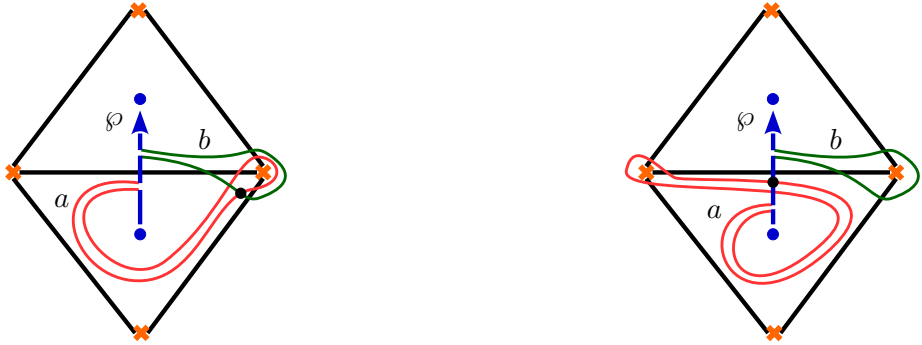
\begin{figure}[tb]
\begin{center}
\begin{overpic}[width=\textwidth]{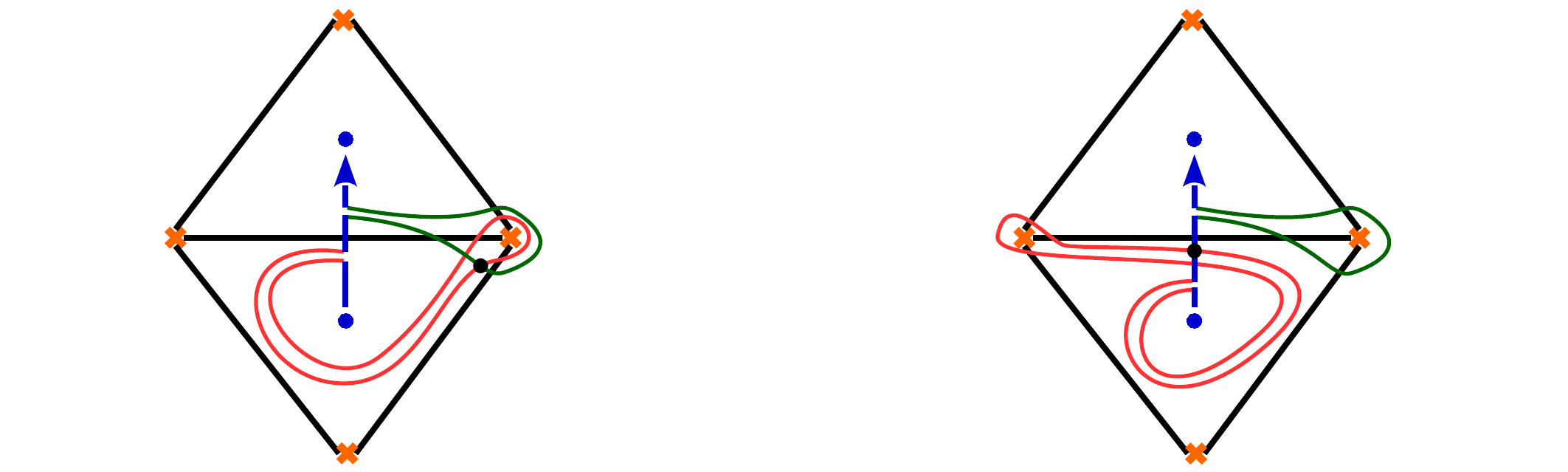}
 	\put (14.7, 12) {$a$}
 	\put (27, 16.9) {$b$}
	 \put (70.5, 11) {$a$}
 	\put (81, 16.9) {$b$}
 	\put (19, 19) {$\wp$}
 	\put (73, 19) {$\wp$}
\end{overpic}
\caption{Two types of intersections of detours (black dots) for the case $m=\bar m = 2$. \emph{Left}: Intersection between detours $a$ and $b$ that circle around the same branch point. \emph{Right}: Intersection due to a full rotation of the detour $a$ around the puncture.}
\label{intersections}
\end{center}
\end{figure}

It would be useful to write down expressions for $Q^\pm_0$ for an edge $\gamma_0$ in a more general BPS graph. A generic situation is shown in Figure~\ref{Qgeneral}, where the edge $\gamma_0$ separates two punctures of~$\cC$, each one surrounded by a cycle of edges, $\gamma_0,\gamma_1, \ldots, \gamma_m$ and $\gamma_0,\gamma_{\bar 1}, \ldots, \gamma_{\bar m}$ (I will assume that the two cycles do not share any edge apart from $\gamma_0$, which will be sufficient for the purpose of this paper). The detours $a,b$ can extend along these cycles, and can even loop around them an infinite number of times. Each time a detour loops fully around a cycle, there is a self-intersection in the concatenated path $ab$ or $ba$. 
Another way for a self-intersection to appear is when the detours $a$ and $b$ circle around the same branch point (Figure~\ref{intersections}).

The definitions~\eqref{Qpmdef} then give (with the notation $X_{0^21} = X_{2\gamma_0 + \gamma_1}$)
\bea
Q_0^- &=& 1 + q^{-\frac12}X_0 + q^{-\frac12}X_{01} +q^{-\frac12} X_{0\bar 1} +q^{-\frac12}X_{01\bar 1} + \cdots + X_{01\cdots m} +  q^{-1} X_{0\bar 1\cdots \bar m} \nn
&& + q^{-\frac12} X_{0^21\cdots m} +  q^{-\frac32} X_{0^2\bar 1\cdots \bar m} +\cdots
+  2q^{-1} X_{0^21\cdots m\bar 1\cdots \bar m}  + \cdots , \nn
Q_0^+ &=& 1 + q^{-\frac12}X_0 + q^{-\frac12}X_{0m} +q^{-\frac12} X_{0\bar m} + \cdots + q^{-1} X_{0m\cdots 1} +  X_{0\bar m\cdots \bar 1} \nn
&&+ q^{-\frac32} X_{0^2m\cdots 1} +  q^{-\frac12} X_{0^2\bar m\cdots \bar 1} + \cdots
+ 2q^{-1} X_{0^2m\cdots 1\bar m\cdots \bar 1} + \cdots ,
\eea
which can be written more conveniently as
\bea\label{Q0pmdenom}
Q_0^- &=& \frac{1+q^{-\frac12}(X_0 + X_{01}+ X_{0\bar 1}+ \cdots + X_{01\cdots m\bar 1\cdots \bar m})+ q^{-1} X_{0^21\cdots m\bar 1\cdots \bar m} }{(1-  X_{01\cdots m})(1- q^{-1}X_{0\bar 1\cdots \bar m})} , \nn
Q_0^+ &=& \frac{1+q^{-\frac12}(X_0 + X_{0m}+ X_{0\bar m}+ \cdots + X_{0m\cdots 1\bar m\cdots \bar 1})+ q^{-1}X_{0^2m\cdots 1\bar m\cdots \bar 1} }{(1- q^{-1}X_{01\cdots m})(1-  X_{0\bar 1\cdots \bar m})} .
\eea
Here the parentheses in the numerators contain all possible combinations of detours that involve each edge at most once, apart from the detours corresponding to closed loops around the punctures such as $\gamma_0 + \gamma_1 + \cdots + \gamma_m$,
which appear in the denominators. These loops do not represent framed BPS states, but rather 2d states with flavor charges living on a surface defect~\cite{Gaiotto:2012rg}. They are therefore not relevant for framed wall-crossing and should be discarded. The remaining numerators can be neatly expressed (with the notation~\eqref{notationShift}) as
\bea
Q_0^- &= & \left(\cQ_0^- \right) ^{-1}_{0^-} \cQ_0^-, \nn
Q_0^+ &= & \cQ_0^+ \left(\cQ_0^+   \right) ^{-1}_{0^-} ,
\eea
with 
\bea \label{cQ0pm}
\cQ_0^- &=& \Phi_{01\cdots m\bar 1\cdots \bar m} \cdots    \Phi_{01\bar 1} \Phi_{0\bar 1} \Phi_{01}\Phi_0  , \nn
\cQ_0^+ &=&  \Phi_0\Phi_{0\bar m} \Phi_{0m} \Phi_{0m\bar m} \cdots \Phi_{0m\cdots 1\bar m\cdots \bar 1} .
\eea
These sequences of quantum dilogarithms again involve all possible combinations of detours without repeated edge, except closed loops around a puncture. 
Note that these expressions are also valid for Riemann surfaces $\cC$ with boundaries.
In this case, some walls of the spectral network end on the boundaries, and the sequence of detours simply terminates when it reaches them.

Comparing with~\eqref{QpmSS} reveals that $\cQ_0^- $ is the part of the quantum spectrum generator $\cS$ that involves $X_0$ after all the $X_0$ have been moved to the right of~$\cS$, while $\cQ_0^+ $ is the part of~$\cS$ with all the $X_0$ to the left of $\cS$: 
\be 
\cS =   \cS_{\cancel 0}\cQ_0^-  = \cQ_0^+ \cS_{\cancel 0}.
\ee
Here $\cS_{\cancel 0}$ is the part of $\cS$ that does not contain $X_0$, or to put it differently, it is $\cS$ with $X_0=0$: $\cS_{\cancel 0} = \cS|_{X_0=0}$. 

In terms of the BPS graph, setting $X_0=0$ can be interpreted as removing the edge $\gamma_0$.
This suggests an iterative procedure: compute $\cQ_0^\pm$ and remove $\gamma_0$, then compute $\cQ_{1,\cancel0}^\pm$ in the simplified BPS graph and remove $\gamma_1$, and so on.
This eventually produces the full quantum spectrum generator
\be
\cS = \cQ_{d,\cancel{0}\cancel{1}\cdots }^- \cdots \cQ_{1,\cancel{0}}^- \cQ_0^- = \cQ_0^+\cQ_{1,\cancel{0}}^+ \cdots \cQ_{d,\cancel{0}\cancel{1}\cdots }^+,
\ee
where $d = \text{rank}\ \Gamma$ is the number of edges in the BPS graph.
This provides a simple iterative method to read off the BPS spectrum from a BPS graph. 
Note that the order in which the edges are removed from the BPS graph corresponds to a choice of BPS chamber, in which the phases of the central charges of BPS states are ordered accordingly.

\section{Examples}

\subsection{Argyres-Douglas theories}

The AD$_k$ theories are associated with a disc $\cC$ with $k+2$ marked points on the boundary~\cite{Gaiotto:2009hg, Gaiotto:2012db, Xie:2012hs}. The determination of the BPS spectrum from the BPS graph is particularly simple since many walls of the spectral network run off to the boundary and do not allow for detours.

\begin{figure}[tb]
\centering
\begin{minipage}{.32\textwidth}
\centering
	\begin{overpic}[width=\textwidth]{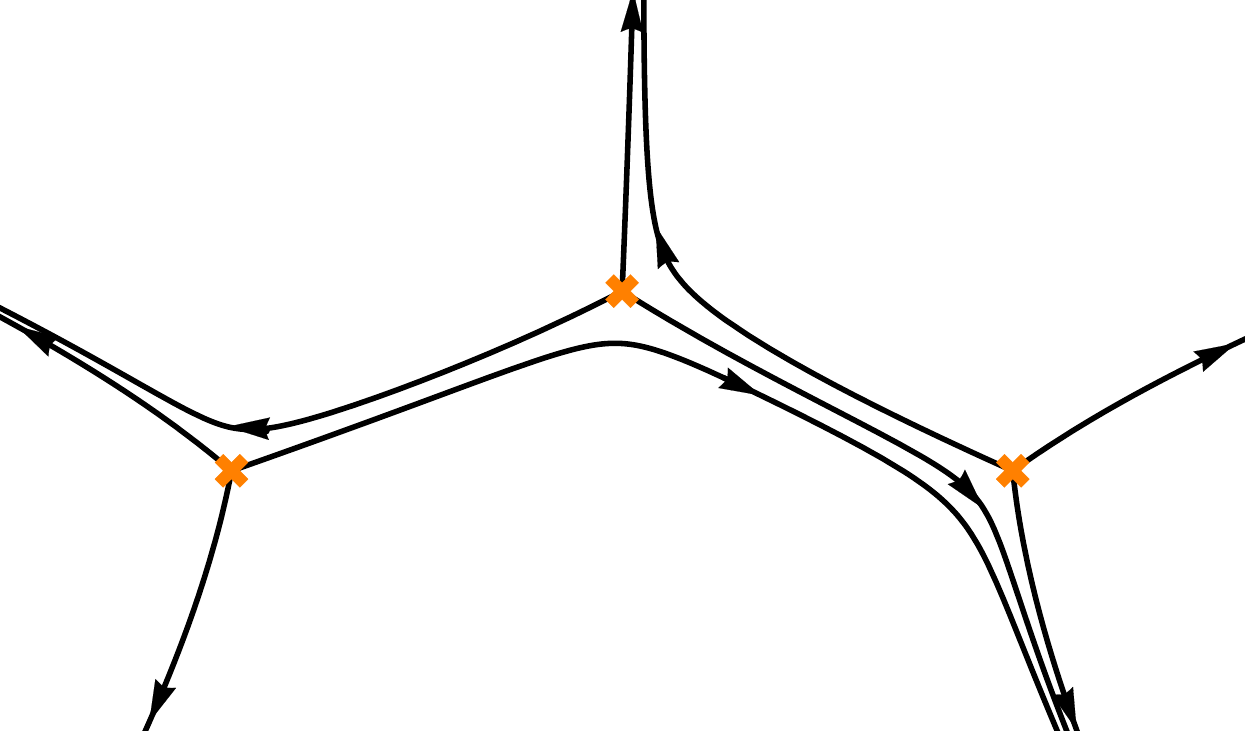}
 		\put (0,55) {$\vartheta_c^-$}
	\end{overpic}
\end{minipage} \hfill
\begin{minipage}{.32\textwidth}
\centering
	\begin{overpic}[width=\textwidth]{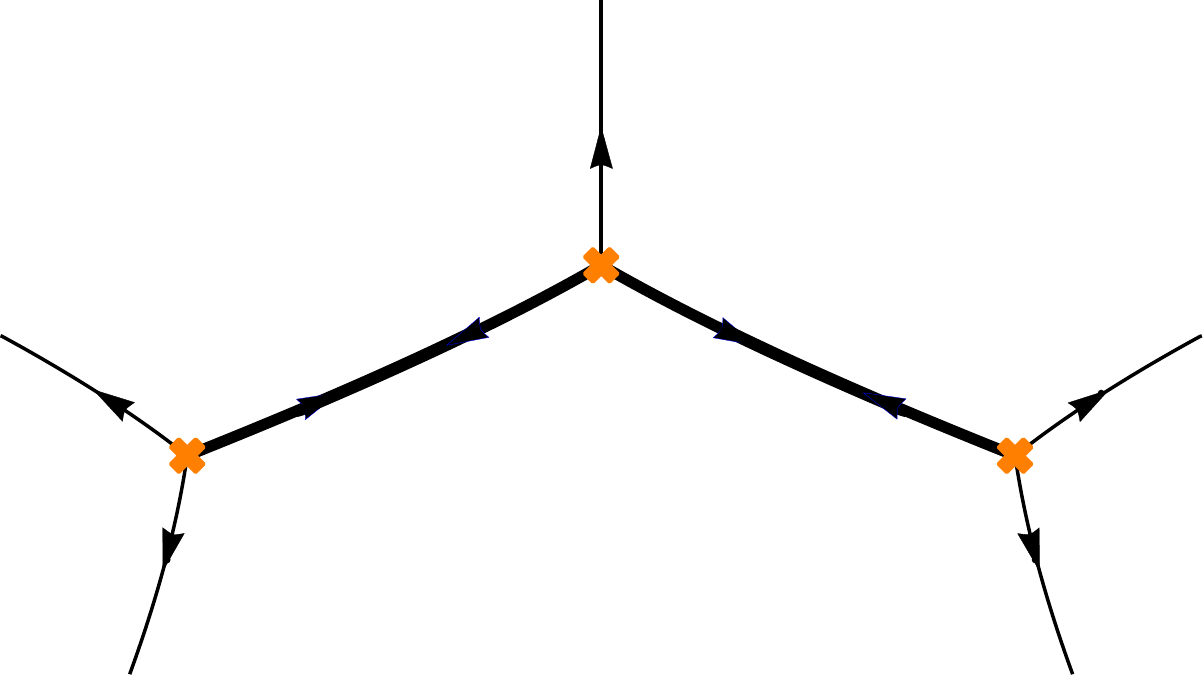}
 		\put (0,55) {$\vartheta_c$}
 		\put (27, 30) {$\gamma_1$}
		\put (65,30) {$\gamma_2$}
	\end{overpic}
\end{minipage}\hfill
\begin{minipage}{.32\textwidth}
\centering
	\begin{overpic}[width=\textwidth]{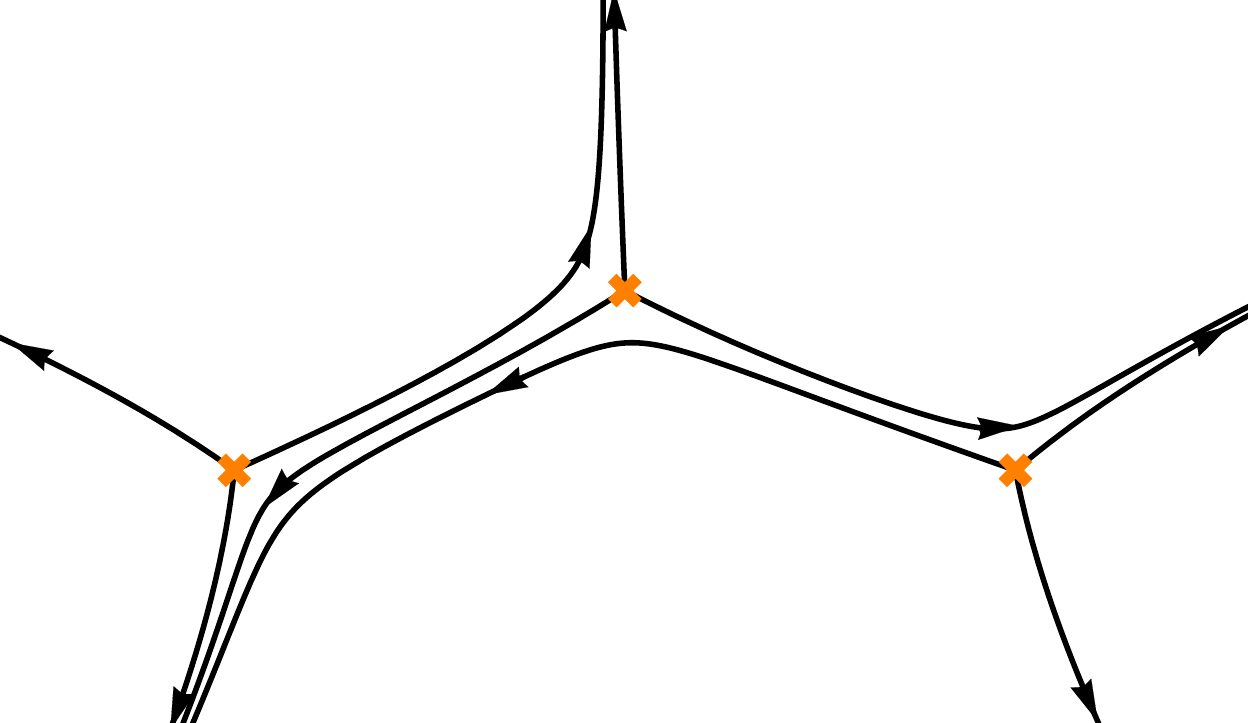}
 		\put (0,55) {$\vartheta_c^+$}
	\end{overpic}
\end{minipage}
\caption{BPS graph for the AD$_3$ theory at $\vartheta_c$, and its American and British resolutions. }
\label{AD3}
\end{figure}

The BPS graph for the AD$_3$ theory has two edges, with $\langle \gamma_1, \gamma_2 \rangle = -1 $ (Figure~\ref{AD3}). 
The $Q^\pm$ are given by
\bea
Q^-_1 &=& 1+ q^{-\frac12} X_{1} = \Phi(q^{-1}X_1)^{-1} \Phi(X_1), \\
Q^+_1 &=& 1+ q^{-\frac12}  (X_{1}+  X_{12})= \Phi(X_1)\Phi(X_{12})\left[\Phi(q^{-1}X_1)\Phi(q^{-1}X_{12})\right]^{-1} , \\
Q^-_2 &=& 1+   q^{-\frac12}(X_{2}+    X_{12}) = \left[\Phi(q^{-1}X_{12})\Phi(q^{-1}X_2)\right]^{-1}\Phi(X_{12})\Phi(X_2), \\
Q^+_2 &=& 1+ q^{-\frac12} X_{2} =\Phi(X_2) \Phi(q^{-1}X_2)^{-1} .
\eea
Depending on the order in which the edges get removed and on the resolution (American or British), the quantum spectrum generator takes one of two forms:
\bea
\cS &=& \cQ_{2,\cancel{1}}^- \cQ_1^- =\cQ_2^+ \cQ_{1,\cancel{2}}^+ = \Phi(X_2) \Phi(X_1), \nn
\cS &=&  \cQ_{1,\cancel{2}}^- \cQ_2^- =\cQ_1^+ \cQ_{2,\cancel{1}}^+ = \Phi(X_1) \Phi(X_{12}) \Phi(X_2).
\eea
The two results are of course related by wall-crossing, via the quantum pentagon identity~\eqref{qPentagon}.

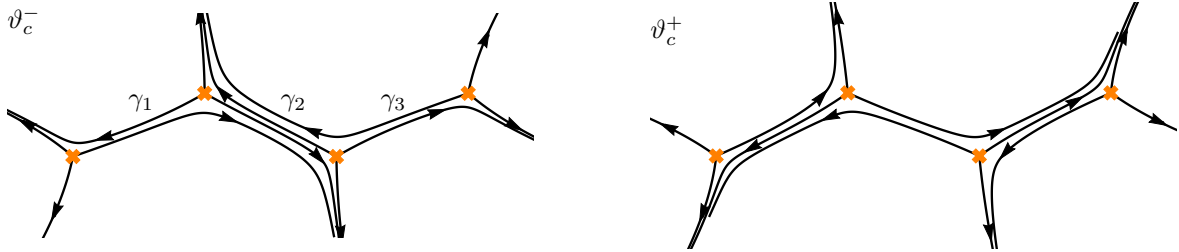
\begin{figure}[tb]
\centering
\begin{minipage}{.45\textwidth}
\centering
	\begin{overpic}[width=\textwidth]{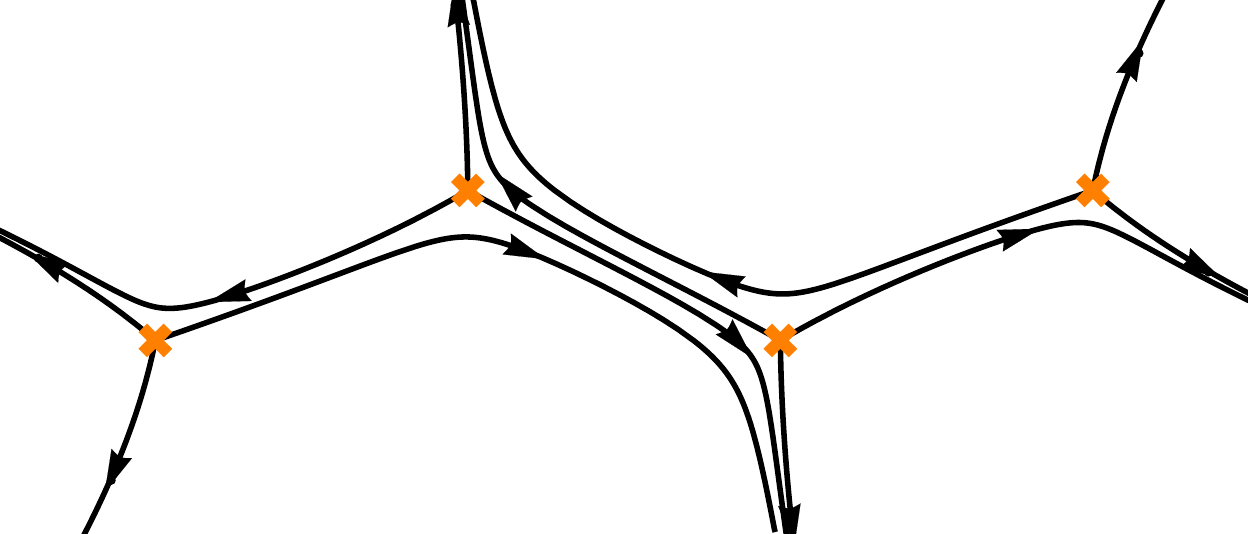}\small
 		\put (0, 40) {$\vartheta_c^-$}
 		\put (23, 25) {$\gamma_1$}
 		\put (52, 25) {$\gamma_2$}
 		\put (71, 25) {$\gamma_3$}
	\end{overpic}
\end{minipage} \hfill
\begin{minipage}{.45\textwidth}
\centering
	\begin{overpic}[width=\textwidth]{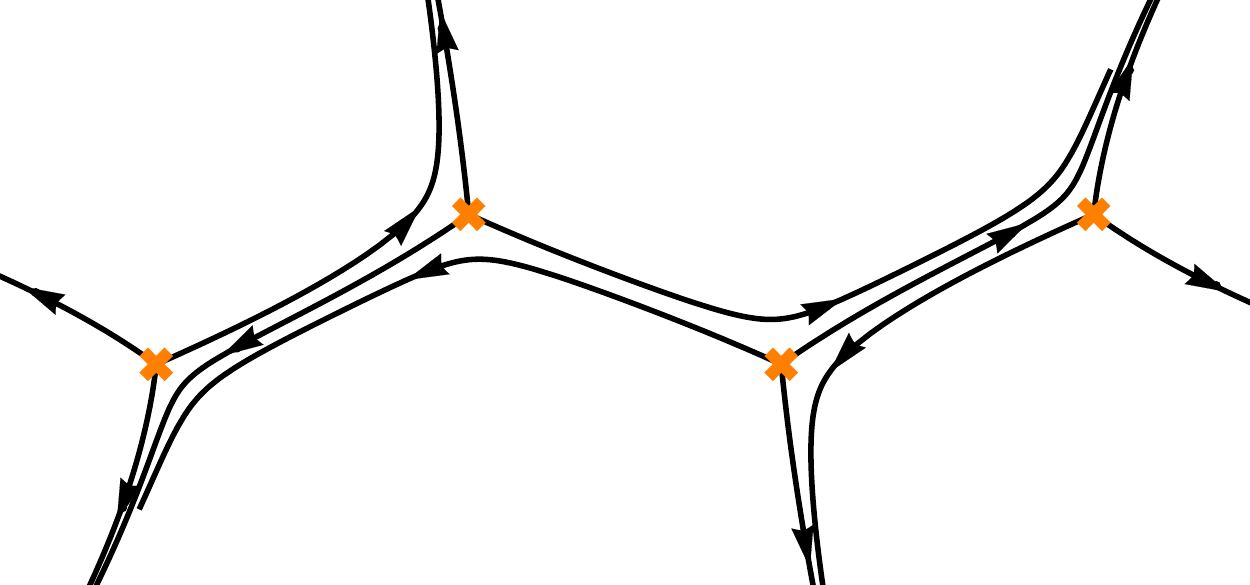}\small
 		\put (0,40) {$\vartheta_c^+$}
	\end{overpic}
\end{minipage}
\caption{American and British resolutions of the BPS graph for the AD$_4$ theory. }
\label{AD4}
\end{figure}

For the AD$_4$ theory there are three cycles, with $\langle \gamma_1, \gamma_2 \rangle = -1 $ and $\langle \gamma_2, \gamma_3 \rangle = 1 $ (Figure~\ref{AD4}). A short-cut to the quantum spectrum generator is to notice that for each edge there is a $Q^\pm$ that is elementary:
\be
Q_1^- =1+q^{-\frac12} X_1  , \quad Q_2^+ =1+q^{-\frac12} X_2, \quad  Q_3^- =1+q^{-\frac12} X_3 .
\ee
Given that $Q^-_{1,3}$ appear for $\vartheta_c^-<\vartheta_c$ and $Q^+_2$ at $\vartheta_c^+>\vartheta_c$, the $X_\alpha$ with $\alpha$ odd should appear on the right of $\cS$, and those with $\alpha$ even on the left. This immediately implies 
\be
\cS  = \Phi(X_2) \Phi(X_1) \Phi(X_3) .
\ee

This approach generalizes to an AD$_k$ theory with arbitrary $k$, where
\bea
Q_\alpha^- &=& 1+q^{-\frac12} X_\alpha   \qqq \text{ for $\alpha$ odd},     \nn
Q_\alpha^+ &=&1+q^{-\frac12} X_\alpha  \qqq \text{ for $\alpha$ even} .
\eea
This implies that the quantum spectrum generator is given by
\be
\cS  = \prod_{\alpha \text{ even}} \Phi(X_\alpha)\prod_{\alpha' \text{ odd}} \Phi(X_{\alpha'}) ,
\ee
in agreement with the BPS spectrum in the ``sink/source chamber'' discovered in~\cite{Cecotti:2010fi}.

\subsection{$T_2$ theory}

\begin{figure}[tb]
\centering
\begin{minipage}{.32\textwidth}
\centering
	\begin{overpic}[width=\textwidth]{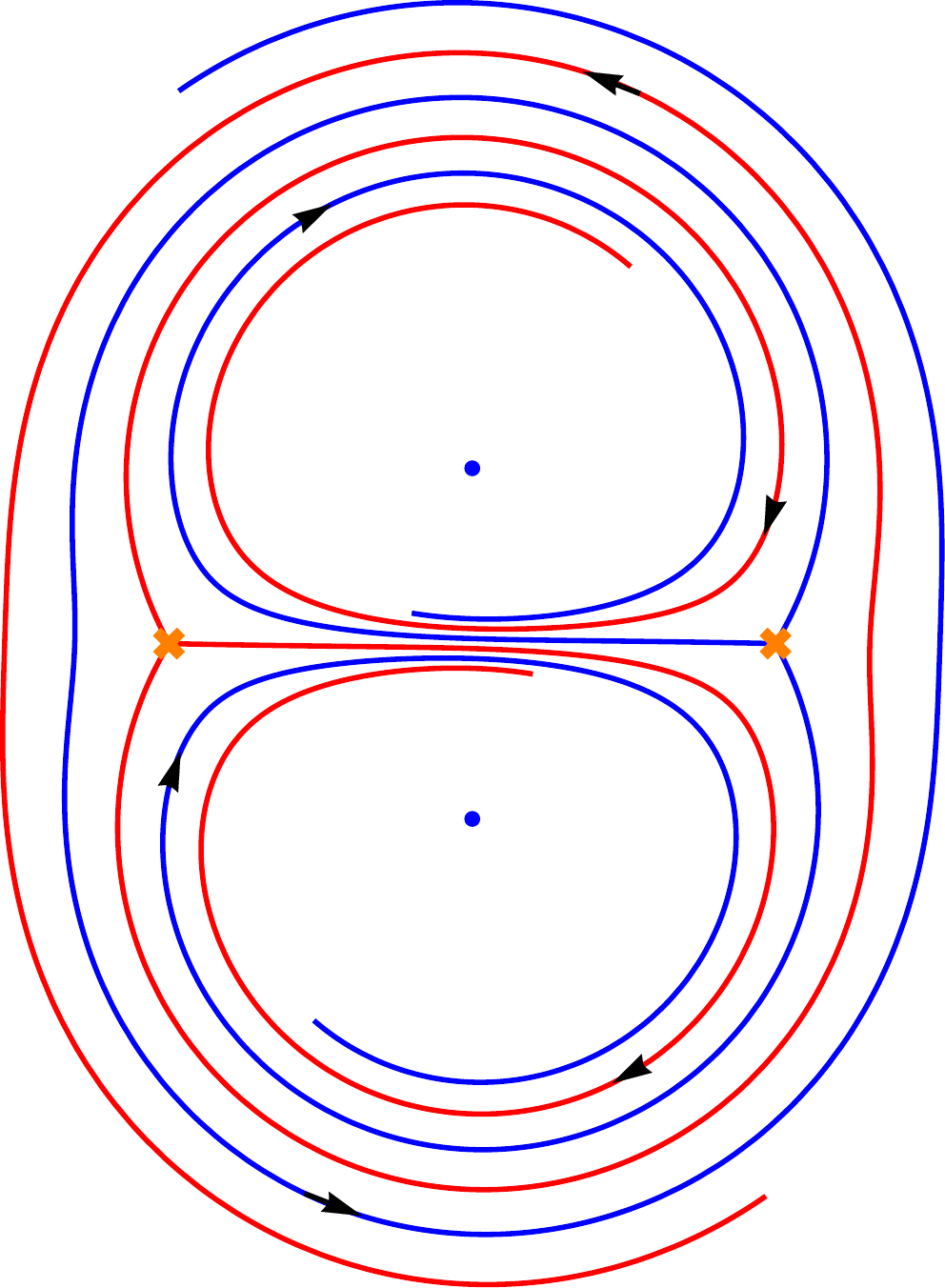}
 		\put (0,100) {$\vartheta_c^-$}
	\end{overpic}
\end{minipage} \hfill
\begin{minipage}{.25\textwidth}
\centering
	\begin{overpic}[width=\textwidth]{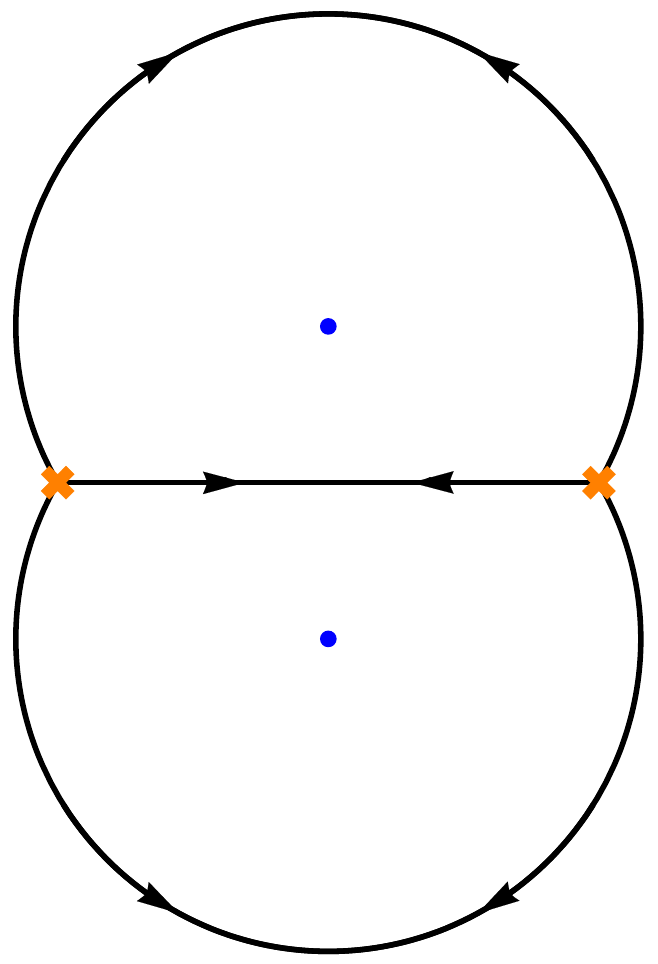}
 		\put (0,110) {$\vartheta_c$}
 		\put (23, 54) {$\gamma_1$}
		\put (23,7) {$\gamma_2$}
		\put (23,101) {$\gamma_3$}
	\end{overpic}
\end{minipage}\hfill
\begin{minipage}{.32\textwidth}
\centering
	\begin{overpic}[width=\textwidth]{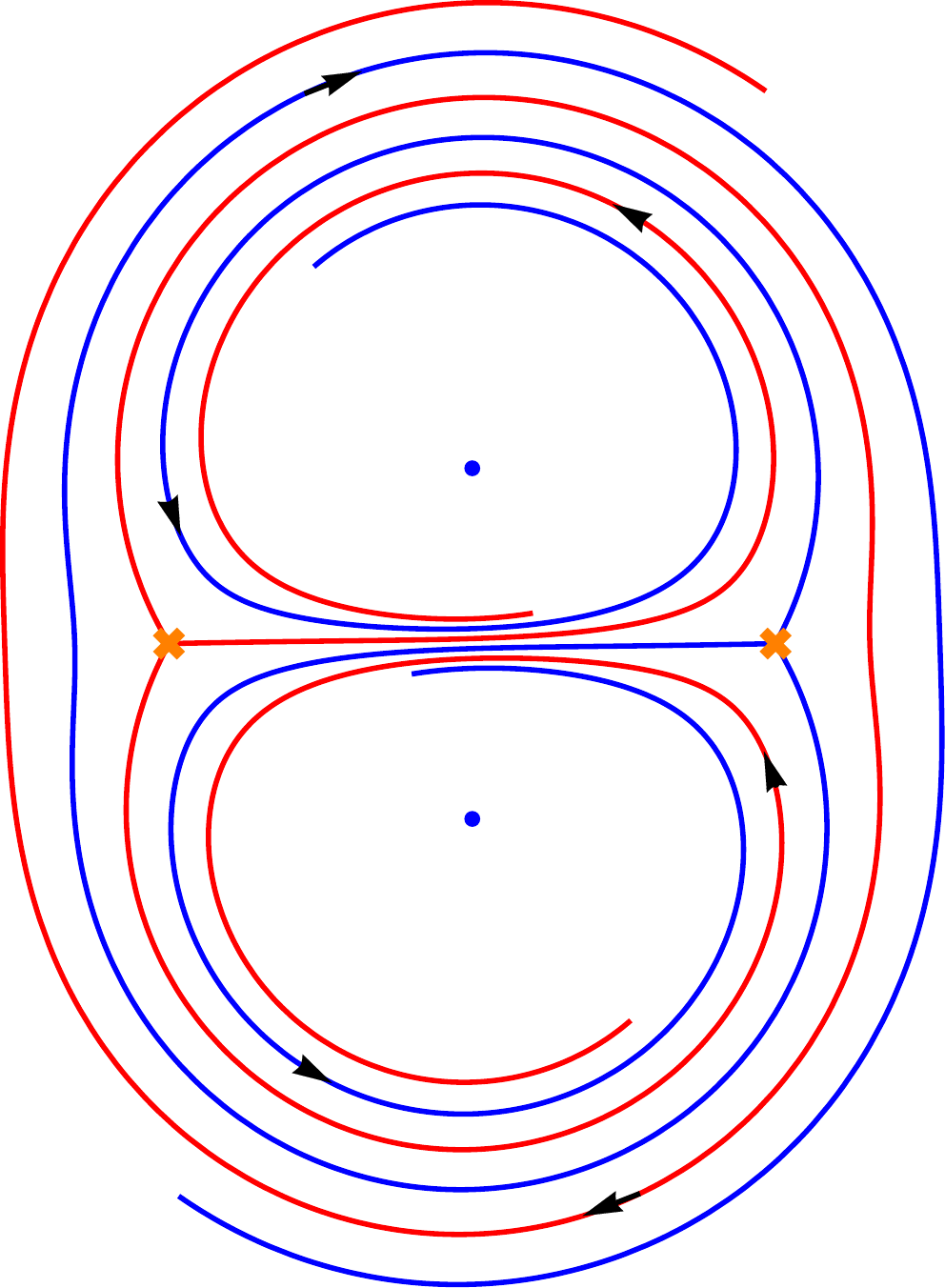}
 		\put (0,100) {$\vartheta_c^+$}
	\end{overpic}
\end{minipage}
\caption{BPS graph for the $T_2$ theory, and its American and British resolutions.}
\label{T2fig}
\end{figure}
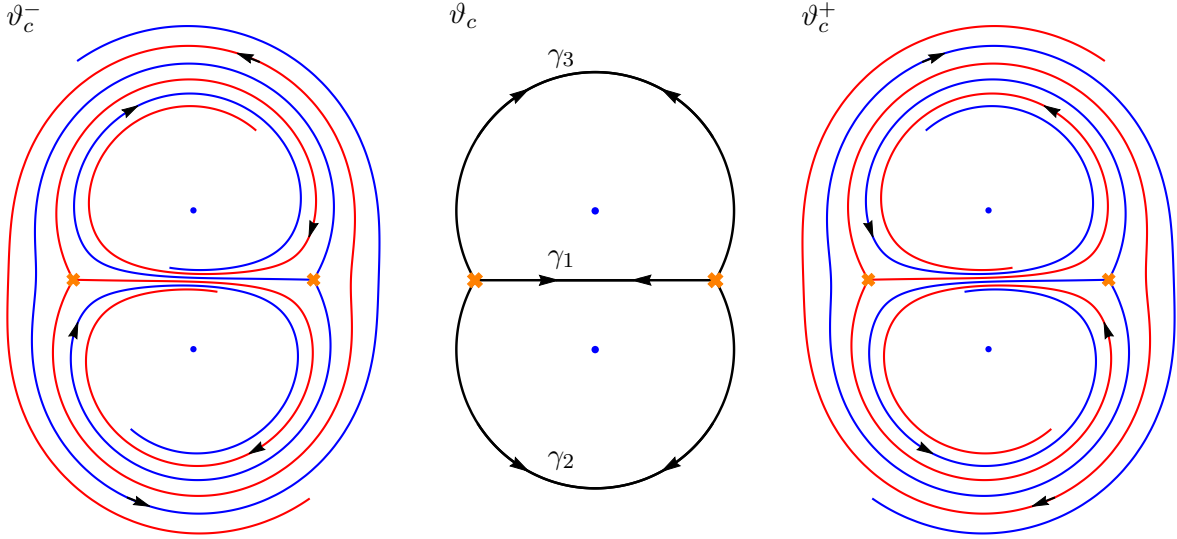

The BPS graph for the $T_2$ theory associated with the three-punctured sphere is shown in Figure~\ref{T2fig}. In this case, the Coulomb branch is trivial and the intersection form vanishes identically. 
Applying~\eqref{Q0pmdenom} for the edge $\gamma_1$ gives
\bea
Q_1^- &=& \frac{1+ q^{-\frac12}(X_1 + X_{123}) + q^{-1} X_{1^223}}{(1-X_{12})(1-q^{-1} X_{13})} , \nn
Q_1^+ &=& \frac{1+ q^{-\frac12}(X_1 + X_{123}) + q^{-1} X_{1^223}}{(1-q^{-1}X_{12})(1- X_{13})} .
\eea
Without the denominators corresponding to 2d flavor states, this can be written as
\be
Q_1^\pm = \Phi(X_1) \Phi(X_{123}) \left[ \Phi(q^{-1}X_1) \Phi(q^{-1}X_{123}) \right]^{-1},
\ee
and so $\cQ_1^\pm  = \Phi(X_1) \Phi(X_{123})$.
The next step is to remove the edge $\gamma_1$ and compute
\be
Q_{2,\cancel 1}^\pm = \Phi(X_2)  \Phi(q^{-1}X_2)^{-1}, \quad Q_{3,\cancel 1}^\pm = \Phi(X_3)  \Phi(q^{-1}X_3)^{-1}.
\ee
This leads to the correct quantum spectrum generator, with four BPS states:
\be 
\cS = \Phi(X_1) \Phi(X_2) \Phi(X_3) \Phi(X_{123}).
\ee

\subsection{Pure $SU(2)$ theory}

\begin{figure}[tb]
\begin{center}
\begin{overpic}[width=0.99\textwidth]{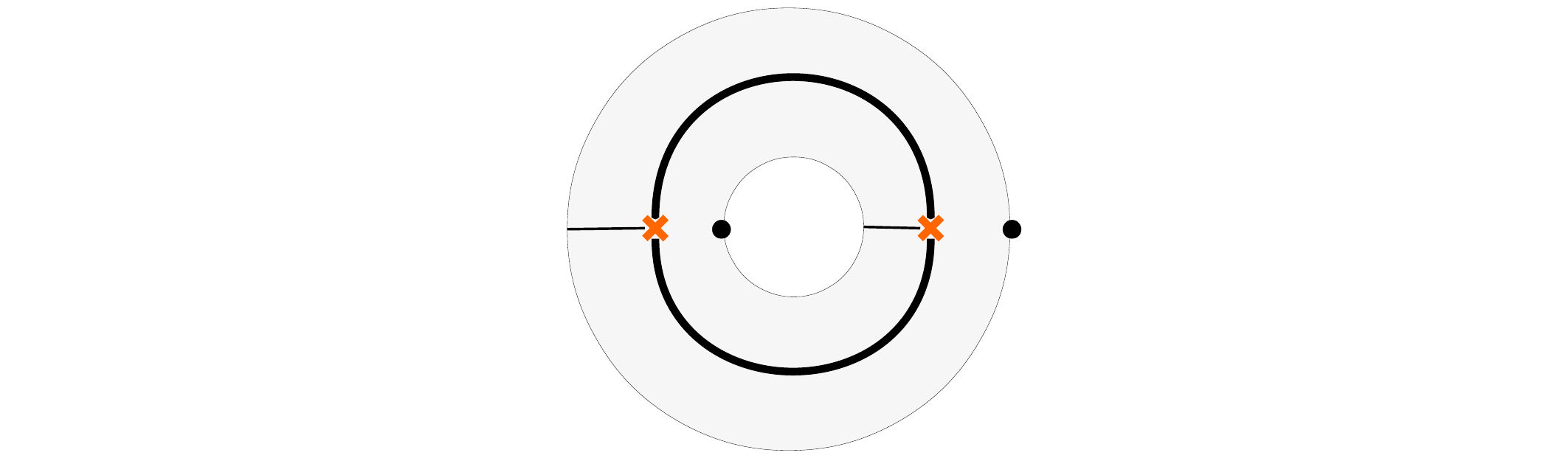}
 	\put (50, 26) {$\gamma_1$}
 	\put (50, 3.5) {$\gamma_2$}
\end{overpic}
\caption{BPS graph for pure $SU(2)$ theory on an annulus with marked points on the boundaries.}
\label{PureSU2}
\end{center}
\end{figure}

Pure $SU(2)$ theory can be realized by taking $\cC$ to be a cylinder with one marked point on each boundary. 
The BPS graph has two edges, with $\langle \gamma_1, \gamma_2\rangle = 2$ (Figure~\ref{PureSU2}).

This shortest way to the quantum spectrum generator is to notice that 
\be 
Q_1^+ = 1 + q^{-\frac12} X_1, \qqq Q_2^- = 1 + q^{-\frac12} X_2,
\ee
which immediately gives to the correct result
\be
\cS = \Phi(X_1) \Phi(X_2).
\ee

As a remark, note that for other choices of resolution the expressions~\eqref{Q0pmdenom} and~\eqref{cQ0pm} are not valid, because some detours $a$ and $b$ can have common edges. Instead we find 
\bea
Q_1^- &=&1+q^{-\frac12}\left(X_1+ [2] X_{12} + X_{12^2} \right)=  \left( \Phi(q^{-1}X_{1})  \Phi(X_2) \right)^{-1} \Phi(X_{1})  \Phi(X_2) , \nn 
 Q_2^+ &=& 1+q^{-\frac12}\left(X_2+ [2] X_{12} + X_{1^22}\right)  =\Phi(X_{1})  \Phi(X_2)  \left( \Phi(X_{1})  \Phi(q^{-1}X_2) \right)^{-1}   ,
\eea
with $[2] = q^{\frac12}+ q^{-\frac12}$. This is however not a big problem, since, as we have just seen, judicious choices of $Q^{\pm}$ are enough to determine the quantum spectrum generator.

\subsection{$SU(2)$ theory with one flavor}

\begin{figure}[tb]
\begin{center}
\begin{overpic}[width=0.99\textwidth]{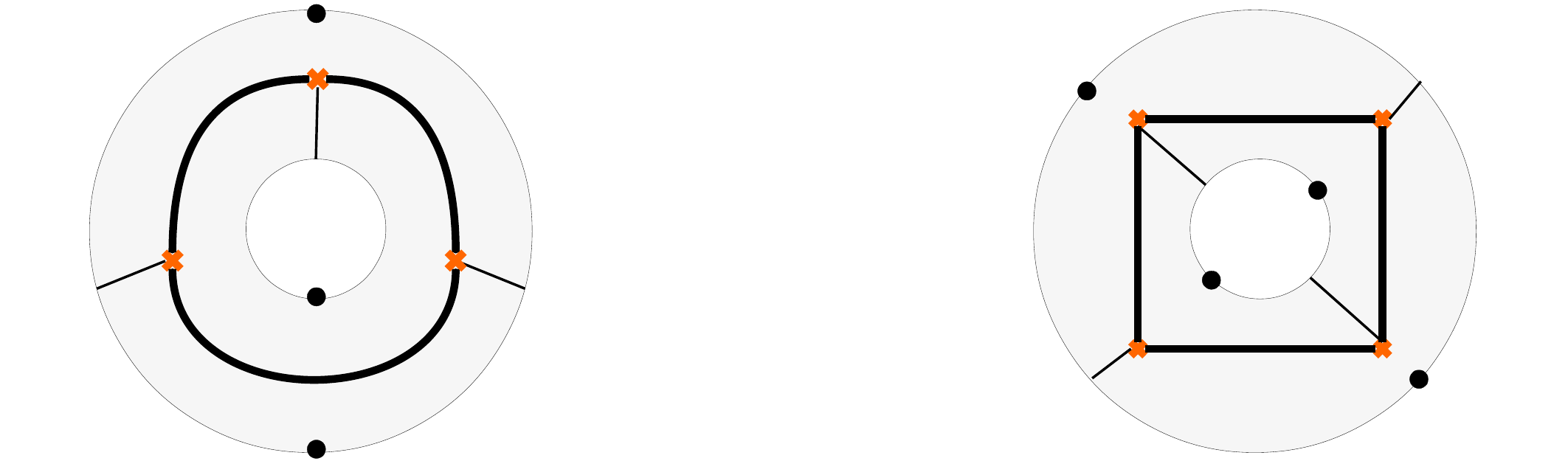}
 	\put (9, 20) {$\gamma_1$}
 	\put (28.7, 20) {$\gamma_2$}
 	\put (19, 6.7) {$\gamma_3$}
 	\put (69.5, 14) {$\gamma_1$}
 	\put (89, 14) {$\gamma_2$}
 	\put (80, 23.5) {$\gamma_3$}
 	\put (80, 5) {$\gamma_4$}
\end{overpic}
\caption{BPS graphs for $SU(2)$ theories with $N_f=1$ (left) and $N_f=2$ (right).}
\label{SU2Nf12}
\end{center}
\end{figure}
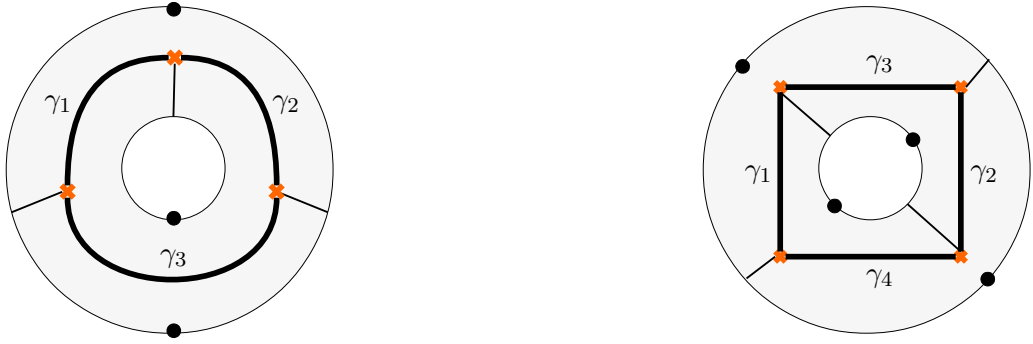

The BPS graph for $SU(2)$ theory with $N_f=1$ flavor is shown on the left of Figure~\ref{SU2Nf12}.
Removing first the edge $\gamma_1$ and then $\gamma_3$ gives
\be
Q_1^+ =1+q^{-\frac12}X_1  ,  \qqq Q_{3,\cancel 1}^-=1+q^{-\frac12}X_3, \qqq   Q_{2,\cancel1 \cancel3}^-  =1+q^{-\frac12}X_2 .
\ee
This leads immediately to the quantum spectrum generator
\be
\cS =  \cQ_1^+\cQ_{3,\cancel{1}}^+ \cQ_{2,\cancel{1}\cancel3}^+= \Phi(X_1) \Phi(X_3) \Phi(X_2).
\ee

\subsection{$SU(2)$ theory with two flavors}

The BPS graph for $SU(2)$ theory with $N_f=2$ flavors is shown on the right of Figure~\ref{SU2Nf12}.
We find 
\begin{align}
Q_1^+ &=1+q^{-\frac12}X_1  ,      &  Q_2^+  &=1+q^{-\frac12}X_2  , \nn 
Q_3^- &=1+q^{-\frac12}X_3  ,      &  Q_4^-  &=1+q^{-\frac12}X_4 ,
\end{align}
which gives
\be
\cS = \Phi(X_1)  \Phi(X_2)\Phi(X_3)\Phi(X_4).
\ee

\subsection{$SU(2)$ theory with three flavors}
 
\begin{figure}[tb]
\begin{center}
\begin{overpic}[width=0.99\textwidth]{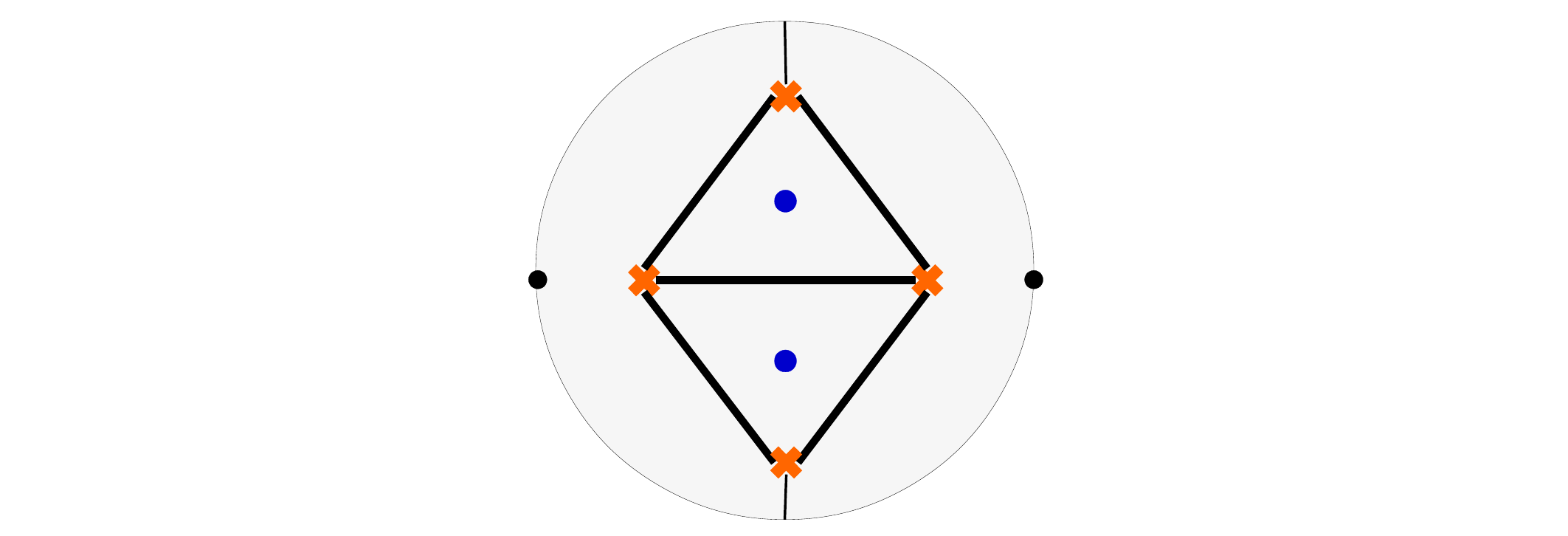}
 	\put (50, 17.5) {$\gamma_1$}
 	\put (41, 10) {$\gamma_2$}
 	\put (57, 10) {$\gamma_3$}
 	\put (57, 21) {$\gamma_4$}
 	\put (41, 21) {$\gamma_5$}
\end{overpic}
\caption{BPS graph for $SU(2)$ theory with $N_f=3$ on a disc with two punctures and two marked points on the boundary.}
\label{SU2Nf3}
\end{center}
\end{figure}
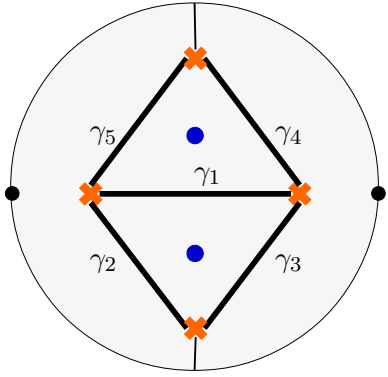

The $SU(2)$ theory with $N_f=3$ flavors can be realized by taking $\cC$ to be a disc with two marked points on the boundary and two punctures. The BPS graph in shown in Figure~\ref{SU2Nf3}.
We start by removing the edges $\gamma_2$ and $\gamma_4$, for which we have (with the notation $\Phi_\alpha = \Phi(X_\alpha)$)
\be
\cQ_2^+ = \Phi_2\Phi_{12} , \qqq \cQ_4^+ = \Phi_4\Phi_{14} .
\ee
It is then easy to compute 
\be
\cQ_{3,\cancel2 \cancel4}^+ = \Phi_3 , \qqq \cQ_{5,\cancel2 \cancel4}^+ = \Phi_5 ,
\ee
and we are finally only left with the edges $\gamma_1$:
\be
\cQ_{1,\cancel2 \cancel4\cancel3 \cancel5}^+ = \Phi_1.
\ee
The quantum spectrum generator is thus given by
\be
\cS = \Phi_{2}\Phi_{12}\Phi_{4} \Phi_{14}\Phi_{3}\Phi_{5}\Phi_{1}.
\ee

We could have equally well started with another edge, say $\gamma_1$, for which we have
\be
\cQ_1^-=\Phi_{12345}\Phi_{1245}\Phi_{1234}\Phi_{124}\Phi_{14}\Phi_{12}\Phi_{1} .
\ee
Then we would find
\be
\cQ_{3,\cancel1}^- = \Phi_3 , \qqq \cQ_{5,\cancel1}^- = \Phi_5 , \qqq \cQ_{2,\cancel1\cancel3\cancel5}^- = \Phi_2 ,  \qqq \cQ_{4,\cancel1\cancel3\cancel5}^- = \Phi_4 ,
\ee
and end up with
\be
\cS = \Phi_{2}\Phi_{4}\Phi_{3} \Phi_{5} \Phi_{12345}\Phi_{1245}\Phi_{1234}\Phi_{124}\Phi_{14}\Phi_{12}\Phi_{1} .
\ee
It is straightforward to check that this is equivalent to the previous result, upon multiple applications of the quantum pentagon identity~\eqref{qPentagon}.

\subsection{$SU(2)$ theory with four flavors}

\begin{figure}[tb]
\begin{center}
\begin{overpic}[width=\textwidth]{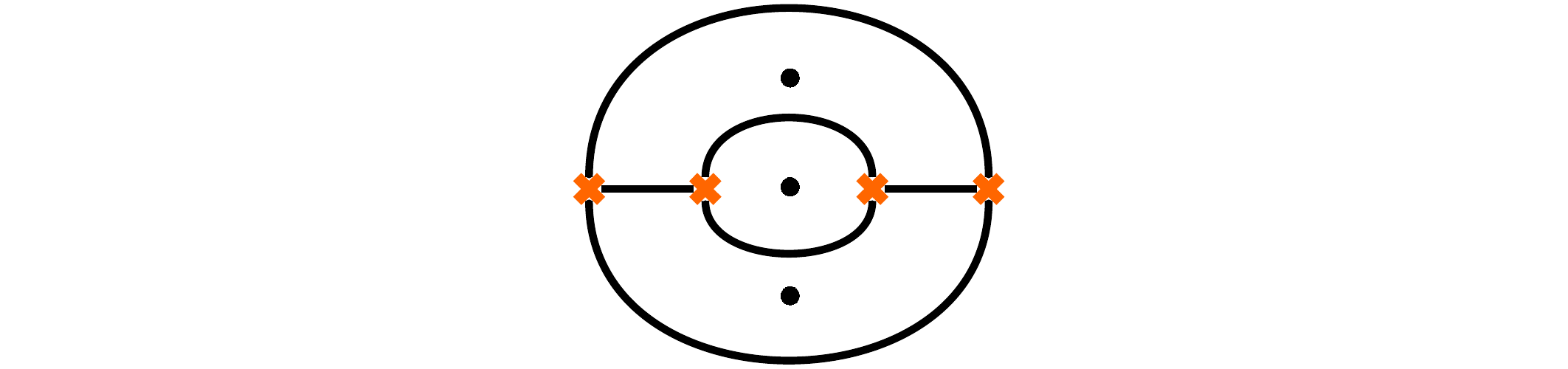}
 	\put (40,13) {$\gamma_1$}
 	\put (58,13) {$\gamma_2$}
 	\put (52,24) {$\gamma_3$}
 	\put (52,17) {$\gamma_4$}
 	\put (52,9) {$\gamma_5$}
 	\put (52,2) {$\gamma_6$}
\end{overpic}
\caption{BPS graph for $SU(2)$ theory with $N_f=4$, on a sphere with four punctures (one of which is at infinity). }
\label{SU2Nf4}
\end{center}
\end{figure}
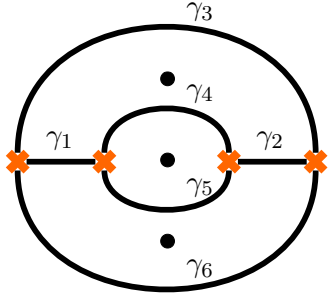

For $SU(2)$ theory with $N_f=4$ flavors, the Riemann surface $\cC$ is a sphere with four punctures. 
The BPS graph that is dual to the BPS quiver used for example in~\cite{Cordova:2016uwk} is shown in Figure~\ref{SU2Nf4}. 
Let's start with the edge $\gamma_3$, for which we find
\be
\cQ_3^+ =  \Phi_3 \Phi_{23} \Phi_{234} \Phi_{236} \Phi_{2346} \Phi_{12346} .
\ee
We then remove the edge $\gamma_3$ and focus on $\gamma_4$. Repeating the process, we end up with 
\bea
&\cQ_{4,\cancel3}^+ =  \Phi_4 \Phi_{14} \Phi_{145} , \qqq 
\cQ_{5,\cancel3\cancel4}^+ =  \Phi_5 \Phi_{25} \Phi_{256} , \qqq 
\cQ_{6,\cancel3\cancel4\cancel5}^+ =  \Phi_6 \Phi_{16} , & \nn
&\cQ_{1,\cancel3\cancel4\cancel5\cancel6}^+ =  \Phi_1 , \qqq\qqq 
\cQ_{2,\cancel3\cancel4\cancel5\cancel6}^+ =  \Phi_2 . & 
\eea
The quantum spectrum generator is thus given by
\be
\cS = \Phi_3 \Phi_{23} \Phi_{234} \Phi_{236} \Phi_{2346} \Phi_{12346}  \Phi_4 \Phi_{14} \Phi_{145}    \Phi_5 \Phi_{25} \Phi_{256}   \Phi_6 \Phi_{16}   \Phi_1  \Phi_2
\ee
Thanks to the quantum pentagon identity~\eqref{qPentagon} we can bring this to the form
\be
\cS = \Phi_6 \Phi_5 \Phi_4 \Phi_3 \Phi_{235} \Phi_{146} \Phi_{25} \Phi_{23} \Phi_{16} \Phi_{14} \Phi_2 \Phi_1,
\ee
which is indeed the result quoted in~\cite{Cordova:2016uwk} (see~\cite{Gaiotto:2009hg, Cecotti:2011gu, Alim:2011kw}). 

\

\begin{figure}
\centering
\begin{minipage}{.45\textwidth}
\centering
\begin{overpic}[width=\textwidth]{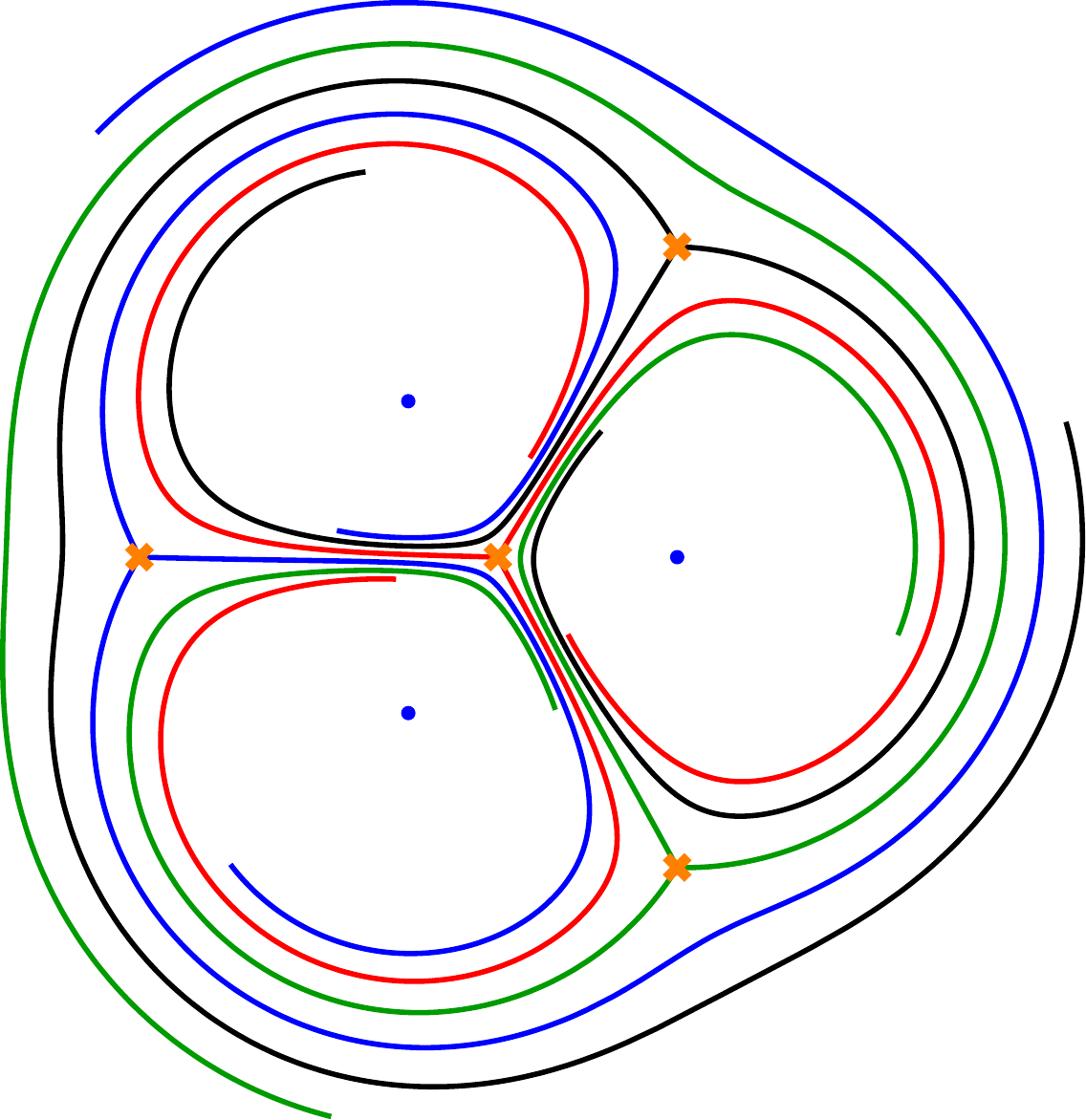}
\end{overpic}
\end{minipage} \hfill
\begin{minipage}{.46\textwidth}
\centering
\begin{overpic}[width=\textwidth]{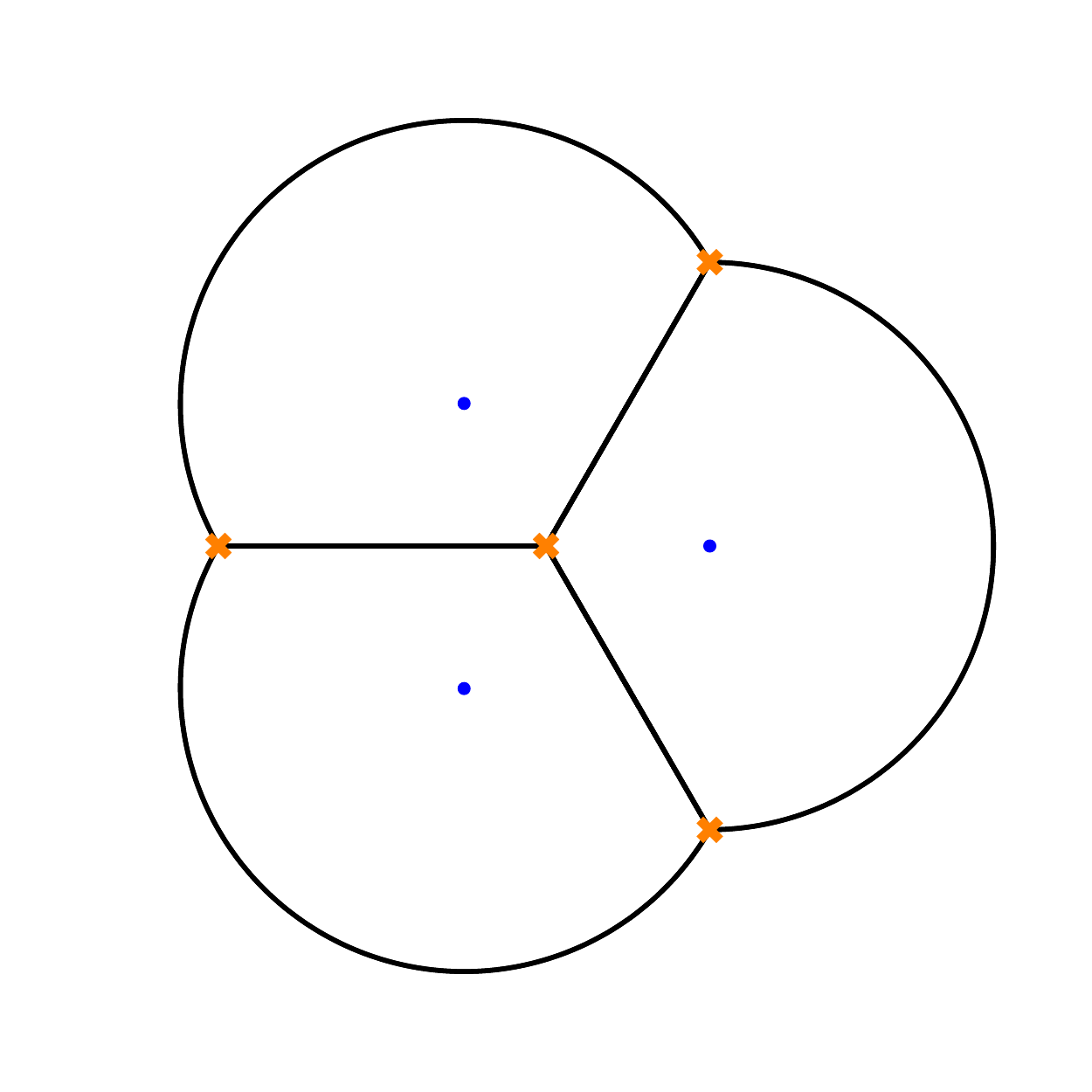}
 	\put (35, 52.5) {$\gamma_1$}
 	\put (31, 81) {$\gamma_2$}
 	\put (31, 16) {$\gamma_3$}
 	\put (60, 36) {$\gamma_5$}
 	\put (60, 60) {$\gamma_4$}
 	\put (85, 48) {$\gamma_6$}
\end{overpic}
\end{minipage}
\caption{Another BPS graph for $SU(2)$ theory with $N_f=4$ flavors. }
\label{SU2Nf4sym}
\end{figure}
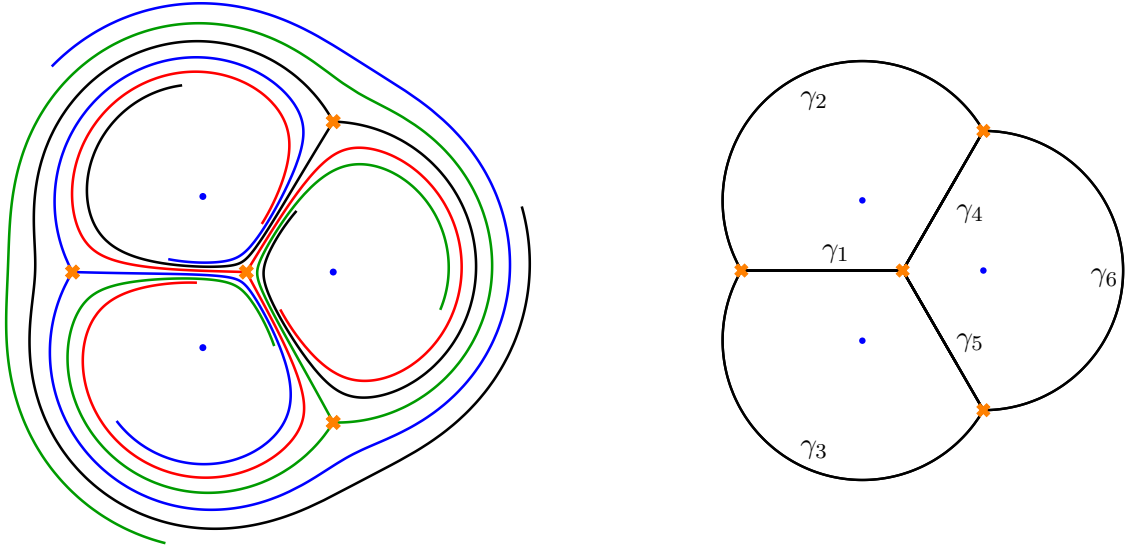

We could also use a different BPS graph, obtained by a flip move~\cite{Gabella:2017hpz} (dual to a mutation of the BPS quiver), see Figure~\ref{SU2Nf4sym}.
We get for example
\bea
&\cQ_1^+  = \Phi_{1} \Phi_{12} \Phi_{15} \Phi_{125} \Phi_{1235} \Phi_{1245} \Phi_{12345} , \qqq  \cQ_{3,\cancel1}^+  = \Phi_{3} \Phi_{36} , \qqq 
\cQ_{4,\cancel1}^+  = \Phi_{4} \Phi_{46} ,   &\nn
&
\cQ_{2,\cancel1\cancel3\cancel4}^+  = \Phi_{2}   , \qqq 
\cQ_{5,\cancel1\cancel3\cancel4}^+  = \Phi_{5}  , \qqq 
\cQ_{6,\cancel1\cancel3\cancel4\cancel2\cancel5}^+  = \Phi_{6} ,  &
\eea
which also gives, after some reordering, a spectrum with 12 BPS states:
\be
\cS =  \Phi_{1} \Phi_{12} \Phi_{15} \Phi_{3} \Phi_{125} \Phi_{1245}\Phi_{36}   \Phi_{4} \Phi_{46}   \Phi_{2}   \Phi_{5}   \Phi_{6} .
\ee

\acknowledgments
I thank Francesco Benini, Clay C\'ordova, Noppadol Mekareeya, and  Shu-Heng Shao for enlightening discussions.
My work was supported by the Swiss National Science Foundation (project P300P2-158440) and by the MIUR-SIR grant RBSI1471GJ,
``Quantum Field Theories at Strong Coupling: Exact Computations
and Applications.''

\bibliographystyle{JHEP}

\providecommand{\href}[2]{#2}\begingroup\raggedright\endgroup

\end{document}